\documentclass[journal,twoside,web]{ieeecolor}
\usepackage{tmi}
\usepackage{bm}
\usepackage{cite}
\usepackage{float}
\usepackage{booktabs}
\usepackage{amsmath,amssymb,amsfonts}
\usepackage{hyperref}
\hypersetup{colorlinks=true,linkcolor=black}
\usepackage{algorithmic}
\usepackage{graphicx}
\usepackage{textcomp}
\usepackage{multirow}

\def\BibTeX{{\rm B\kern-.05em{\sc i\kern-.025em b}\kern-.08em
    T\kern-.1667em\lower.7ex\hbox{E}\kern-.125emX}}
\begin{document}
\title{MOdel-based SyntheTic Data-driven Learning (MOST-DL): Application in Single-shot  T$_2$ Mapping with Severe Head Motion Using Overlapping-echo Acquisition
}
\author{Qinqin Yang, Yanhong Lin, Jiechao Wang, Jianfeng Bao, Xiaoyin Wang, Lingceng Ma, Zihan Zhou,\\ Qizhi Yang, Shuhui Cai, Hongjian He, Congbo Cai, Jiyang Dong, Jingliang Cheng,\\ Zhong Chen, Jianhui Zhong
\thanks{This work was supported by the National Natural Science Foundation of China under grant numbers 82071913, 11775184, U1805261 and 81671674, and Leading (Key) Project of Fujian Province 2019Y0001. }
\thanks{Corresponding authors: Congbo Cai (e-mail: cbcai@xmu.edu.cn) and Jiyang Dong (e-mail: jydong@xmu.edu.cn). }
\thanks{Qinqin Yang, Yanhong Lin, Jiechao Wang, Lingceng Ma, Qizhi Yang, Shuhui Cai, Zhong Chen, Congbo Cai and Jiyang Dong are with the Department of Electronic Science, Xiamen University, Xiamen, 361005, China. }
\thanks{Jianfeng Bao and Jingliang Cheng are with the Department of Magnetic Resonance Imaging, The First Affiliated Hospital of Zhengzhou University, Zhengzhou University, Zhengzhou, 450052, China.}
\thanks{Xiaoyin Wang, Zihan Zhou and Hongjian He are with the Center for Brain Imaging Science and Technology, College of Biomedical Engineering and Instrumental Science, Zhejiang University, Hangzhou, Zhejiang, 310027, China.}
\thanks{Jianhui Zhong is with the Department of Imaging Sciences, University of Rochester, Rochester, New York, USA.}
}
\maketitle

\begin{abstract}
Use of synthetic data has provided a potential solution for addressing unavailable or insufficient training samples in deep learning-based magnetic resonance imaging (MRI). However, the challenge brought by domain gap between synthetic and real data is usually encountered, especially under complex experimental conditions. In this study, by combining Bloch simulation and general MRI models, we propose a framework for addressing the lack of training data in supervised learning scenarios, termed MOST-DL. A challenging application is demonstrated to verify the proposed framework and achieve motion-robust T$_2$ mapping using single-shot overlapping-echo acquisition. We decompose the process into two main steps: (1) calibrationless parallel reconstruction for ultra-fast pulse sequence and (2) intra-shot motion correction for T$_2$ mapping. To bridge the domain gap, realistic textures from a public database and various imperfection simulations were explored. The neural network was first trained with pure synthetic data and then evaluated with \emph{in vivo} human brain. Both simulation and \emph{in vivo} experiments show that the MOST-DL method significantly reduces ghosting and motion artifacts in T$_2$ maps in the presence of unpredictable subject movement and has the potential to be applied to motion-prone patients in the clinic. Our code is available at \url{https://github.com/qinqinyang/MOST-DL}.
\end{abstract}


\section{Introduction}
\label{sec:introduction}
\IEEEPARstart{D}{ata} , algorithms and computing power are the troika of modern artificial intelligence (AI) \cite{b1}. As the first step in AI-based medical imaging processing, many problems come down to insufficient or imperfect data, especially in magnetic resonance imaging (MRI) due to the significant economic burden and long acquisition time for data collection \cite{b2}. In the last decade, many AI-based methods have achieved excellent results in one or a few public datasets but faced challenges in translating into broad clinical applications due to differences among various experimental instruments and situations. Collecting raw data in medical imaging is relatively easy, while data labeling (e.g., informative annotations) is expertise-dependent and often prohibitively time-consuming. Furthermore, training labels may not be available for some complex situations, such as the difficulty of measuring quantitative physical parameters or the irreversibility of the behavior during data collection.

With the development of computer-aided simulation and high-quality rendering technology, synthetic data is increasingly used in AI systems \cite{b3}. In medical imaging, synthetic data has drawn significant attention and been used to address the lack of large datasets \cite{b4} and has provided powerful solutions in applications such as cardiac imaging \cite{b5,b6,b66} and nuclei segmentation in histopathologic images \cite{b7}. Among these, the data-driven (model-free) algorithms, especially generative adversarial networks (GANs), play a key role in generating realistic synthetic data. Learning in synthetic data could accelerate the rollout of data-driven learning algorithms through lower-cost and faster data collection. Furthermore, synthetic data can protect patient privacy and enable greater reproducibility in research. Despite many advantages, data-driven synthesis methods are constrained by the size of the available training dataset, and the biased datasets may lead the trained model towards overrepresented conditions. Chen \emph{et al.} \cite{b8} have expressed concerns about the proliferation of synthetic data created by data-driven methods and recommended the use of simulation-based synthetic data created from forward models \cite{b9} (e.g., existing clinical reference standards, medical prior knowledge and physical laws), which may have regulatory advantages and better interpretability.

Patient motion during MRI scan results in phase mismatch and image artifacts, which can degrade image quality, especially in quantitative MRI (qMRI). Recently, an increasing number of algorithms involving motion correction in qMRI were proposed, and most of them focus on multi-shot sequences \cite{b10,b11,b12}. Due to the irreversible nature of motion, the single-shot acquisition is in general more robust to subject motion (especially severe motion) compared with the multi-shot acquisition. Multiple overlapping-echo detachment (MOLED) sequence \cite{boled1,boled2,boled3}, proposed by our group, has been successfully applied in single-shot qMRI with high accuracy. In MOLED acquisition, overlapping-echo signals containing different phase evolution and parameter weighting are encoded and collected in a single scan with echo planar imaging (EPI) readout. To reconstruct quantitative map from overlapping-echo signals, traditional numerical optimization method was initially used for signal separation \cite{boled1,boled3} but was subsequently replaced by synthetic data-driven learning method based on convolutional neural network (CNN) \cite{boled4}. Though the original MOLED studies demonstrated the robustness to motion \cite{boled1}, the effect of severe motion has not been thoroughly investigated, especially when the parallel imaging technique is applied. Futhermore, synthetic data-driven methods will face greater challenges in complex situations (i.e., parallel imaging and subject motion), and the synthetic-to-real domain gap needs to be further bridged.

In this work, we combine Bloch simulation and general MRI models to generate synthetic training samples for deep learning motion-corrupted parallel MRI, termed MOdel-based SyntheTic Data-driven Learning (MOST-DL). To further bridge the synthetic-to-real domain gap, real anatomical textures from public database were exploited, along with simulation and randomization of non-ideal experimental conditions. With the help of MOST-DL, a challenging application was realized, i.e., T$_2$ mapping under severe head motion for special subjects such as the elderly, children and patients suffering from stroke, emergency trauma, psychological disorders and epilepsy in clinical practice. The single-shot MOLED sequence was applied to acquire signals with different TE weighting at high efficiency, together with the parallel imaging technique to reduce image distortion. The process can be separated into two independent sub-tasks, i.e., (1) parallel reconstruction for ultra-fast sequence and (2) end-to-end T$_2$ mapping with intra-shot motion correction, both of which suffer from difficulty in ‘ground truth’ acquisition.

\section{Related Works}
\subsection{Parallel Reconstruction for Ultra-fast Pulse Sequence}
In the field of EPI/MOLED acquisition, parallel imaging is applied to reduce distortions from B$_0$ inhomogeneity and lessen T$_2$ blurring instead of acquisition acceleration \cite{bepigrappa,bepi1}. The autocalibration signal (ACS) used for interpolation kernel estimation is acquired prior to the under-sampled data, resulting in additional scan time and increased sensitivity to subject motion. Therefore, high-performance and robust calibrationless parallel reconstruction is increasingly becoming a vital factor in under-sampling EPI/MOLED acquisition. 

Shin \emph{et al.} \cite{bepi2} are one of the first to achieve calibrationless parallel imaging reconstruction. They proposed simultaneous autocalibrating and k-space estimation (SAKE) method, which formulates parallel reconstruction as low-rank matrix completion utilizing the redundancy from multi-coil k-space. Similarly, Lee \emph{et al.} \cite{bepi3} proposed an annihilating filter-based low-rank Hankel matrix completion, termed ALOHA algorithm, to perform Nyquist ghost correction and parallel reconstruction in EPI acquisition. However, the low-rank matrix-based methods suffer from high computational costs and often fail to remove the artifacts in under-sampled EPI data due to the uniform Cartesian sampling \cite{bepi4}. Inspired by ALOHA algorithm, Lee \emph{et al.} \cite{bepi5} further improved the result by using a deep neural network. Though the deep learning method has already achieved calibrationless reconstruction of EPI data, it still needs a large number of ALOHA reconstructed images as labels, which may introduce additional reconstruction error in network training and is inefficient as the authors reported. Therefore, data synthesis has the potential to provide a large number of high-quality training samples for calibrationless parallel reconstruction, especially for ultra-fast sequences that are difficult to acquire the ground truth.

\subsection{Deep Learning for MRI Motion Correction}
Regarding motion correction in MRI, most existing deep learning approaches are based on motion simulation from real-world motion-free data \cite{bmotionimage,bmotionimage1,bmotionimage2,bmotionimage3,bmotionksp}. Among the state-of-the-art methods, a representative work presented by Johnson \emph{et al.} \cite{bmotionimage} performed motion simulation in motion-free MR images and combined different motion frames in a new k-space to generate motion-corrupted samples. In order to improve the simulation accuracy, Duffy \emph{et al.} \cite{bmotionksp} performed motion simulation by phase shift and rotation in k-space with non-uniform fast Fourier transform (NUFFT). These works involve direct motion operation and interpolation in acquired MR images, which can be called retrospective motion simulation. However, the retrospective approaches still require a large number of real-world motion-free data using specific pulse sequences. They cannot simulate sub-voxel level motion, and ignore the effects caused by RF inhomogeneity and subject motion before the sampling stage (e.g., during diffusion or MOLED encoding).

Additionally, motion correction has always been studied as a separate step, which has a negative impact on qMRI \cite{bmotionqmri}. Although single-shot MRI scan is robust to slight subject motion, some problems still occur under severe motion, especially in qMRI. Therefore, we combine the process of motion correction and relaxation parameter mapping to avoid the secondary propagation of error in a cascade framework.

\subsection{Synthetic Data-driven Learning in MRI}
Quantitative MR parametric mapping is one of the most successful tasks where synthetic data has been applied, such as MR fingerprinting \cite{bmostmrf,bmostmrf2} and chemical-exchange-saturation-transfer (CEST) imaging \cite{bmostcest}. However, these works are limited to voxel-level Bloch simulation, which is difficult to generalize to motion scenario or other imaging sequences. Some previous works proposed by Liu \emph{et al.} \cite{bmostliu1,bmostliu2} also involve synthetic data in dynamic imaging and qMRI. They created discrete numerical phantoms covering various tissue types and assigned the same value of parameters to each type of tissue, which result in excessive smoothing and loss of detailed texture in final templates. Therefore, their synthetic data are only used to verify the proposed algorithms, and a large amount of real data are still required when transformed to the real world. Besides, the estimation and inversion of various electromagnetic parameters benefit from synthetic data-based methods, such as quantitative susceptibility mapping (QSM) \cite{bmostqsm1,bmostqsm2} and electrical properties tomography (EPT) \cite{bmostept}. These methods have achieved high performance in solving specific problems but are difficult to generalize to other applications. 

Previously our group introduced synthetic data in MRI reconstruction based on 2D Bloch equation evolution, which was used in training deep neural networks to achieve end-to-end T$_2$ mapping from MOLED/OLED images \cite{boled2,boled4} and distortion correction in gradient-echo EPI sequence \cite{bmostgre}. However, these synthetic data were initially created by geometrical shapes such as ellipses, triangles and rectangles, which are quite different from anatomical textures and cause some degree of domain gap between synthetic and real images. Moreover, parallel imaging and subject motion as major MRI issues were not considered in the modeling, which limited the generalizability of the initial version.

This paper builds on our previous works, and the contribution and novelty can be summarized as follows:
\begin{itemize}
	\item We combine the Bloch simulation with a multi-coil parallel imaging model and a rigid motion model, which further extends the application scenarios of synthetic training data. Based on this, we apply calibrationless parallel reconstruction for ultra-fast sequence and intra-shot motion correction in qMRI.
	\item Unlike previous works \cite{bmostgre,bmostqsm2,boled2,boled4,bmostmrf,bmostmrf2,bmostcest,bmostept,bmostliu1,bmostliu2}, rich anatomical texture priors from a publicly available database are used as parametric templates instead of geometrical shapes or numerical phantoms, which allows the generation of data closer to the real features;
	\item Various non-ideal factors are considered in this framework. In particular, the subject motion is modeled at sub-voxel level during Bloch simulation. Moreover, non-ideal factors reconstruction is used as a quality control indicator for secondary validation of the reliability in data generation.
\end{itemize}

\begin{figure*}
	\centering
	\includegraphics[scale=.19]{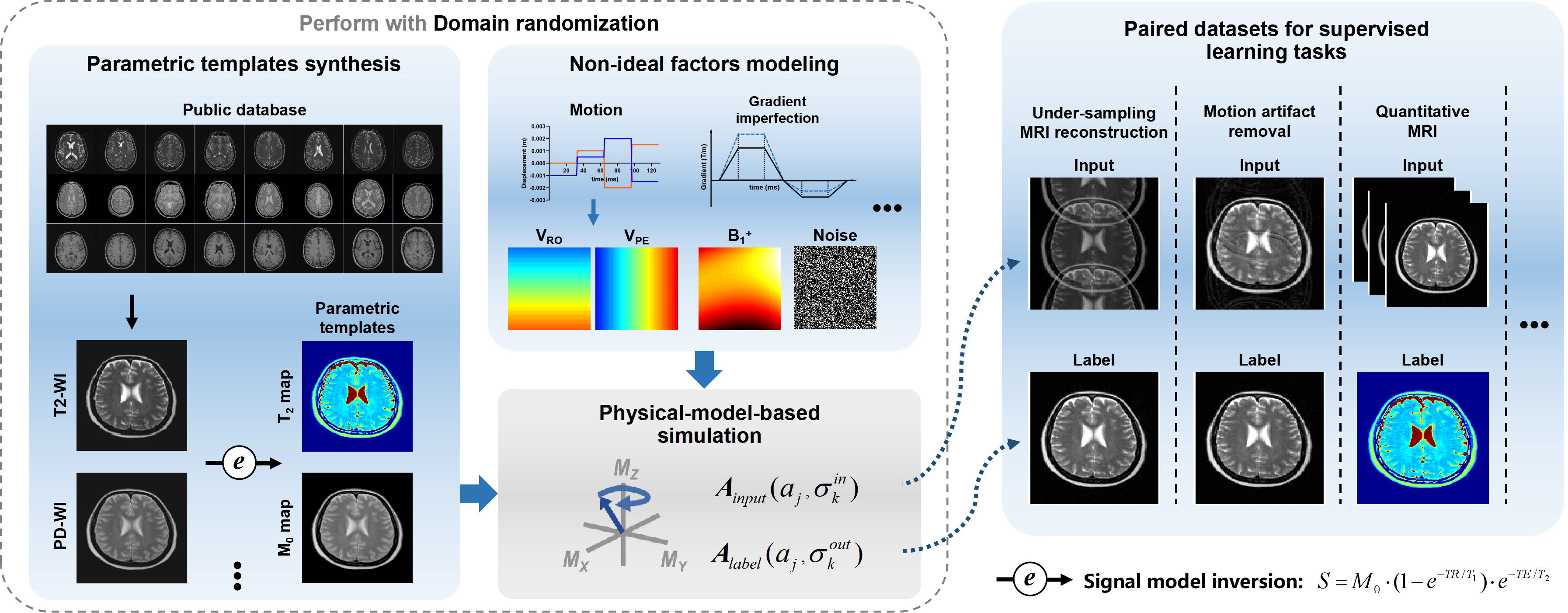}
	\caption{Overview of the MOST-DL framework. Parametric templates synthesis: multi-contrast images from a public database are transformed to parametric templates (distribution $P_{\xi}(a)$) based on the signal model. Non-ideal factors modeling: field inhomogeneity, unexpected motion, noise and instrument imperfections (distribution $P_{\xi}(\sigma)$) are generated by randomization. Physical-model-based simulation: paired datasets for supervised learning tasks are generated by imaging models with Bloch simulation of the task-specific pulse sequence, $A_{input}$ and $A_{label}$ are the forward models to generate the inputs and labels.}
	\label{FIG1}
\end{figure*}

\section{Model-based Synthetic Data-driven Learning}
\subsection{Problem Formulation}
The MRI system can be thought of as a forward physical model $\bm{A}$ that acts on J tissue relaxation parameters $a_j$, which result in measurements $b$ in the image domain. For example, $a_j$ represents T$_1$, T$_2$, and M$_0$ in qMRI. Therefore, the forward of a general imaging problem can be formulated as: 
\begin{equation}b=\bm{A}(a_j,\sigma_k)+\epsilon.\label{eq1}\end{equation}
where $\sigma_k$ denotes K non-ideal factors and $\epsilon$ is the noise in the measured data. The non-ideal factors, $\sigma_k$, consist of field inhomogeneity ($\Delta$B$_0$, B$_1^+$, B$_1^-$), unexpected motion, instrument imperfections, and so on.

Typically, data-driven learning algorithms aim to perform an end-to-end mapping between source data $b_s$ and target data $b_t$ as:
\begin{equation}\hat{b}_{t}=f(b_s;\theta_\Omega).\label{eq2}\end{equation}
where, $f$ is a learning-based model such as a convolutional neural network (CNN), which depends on the trainable parameters $\theta$ of a policy $\Omega$. To solve this domain transformation problem, we can optimize the function:
\begin{equation}\hat{\theta}=\mathop{\arg\min}\limits_{\theta}\mathbb{E}_{(b_s,b_t)\sim P(b)}L[f(b_s;\theta_\Omega)-b_t].\label{eq3}\end{equation}
where $P(b)$ denotes the distribution of measured training sets, and $L[\cdot]$ is the loss function. $\mathbb{E}_{(b_s,b_t)\sim P(b)}[\cdot]$ represents the expectation of loss function when a training sample $(b_s, b_t)$ is drawn from distribution $P(b)$. By incorporating MRI physical operator (1) into Equation (3), the optimization can eliminate the dependency on paired samples, which forms self-supervised learning \cite{bmostliu2}. The optimization can now be formulated as follows:
\begin{equation}
	\begin{aligned}
		& \hat{\theta}=\mathop{\arg\min}\limits_{\theta}\mathbb{E}_{a\sim P(a),\sigma\sim P(\sigma)}\\
		& L[f(\bm{A}_{input}(a_j,\sigma_k^{in});\theta_\Omega)-\bm{A}_{label}(a_j,\sigma_k^{out})].
	\end{aligned}
\end{equation}
here, $P(a)$ and $P(\sigma)$ denote the distribution of tissue relaxation parameters (parametric templates) and non-ideal factors. $\bm{A}_{input}$ and $\bm{A}_{label}$ are the forward models to generate source and target data with the corresponding non-ideal factors $\sigma_k^{in}$ and $\sigma_k^{out}$. Ideally, we would like to apply a model trained on synthetic data to real data. To achieve this purpose, we need to introduce \emph{domain randomization}\cite{bdr1}, making the distribution of synthetic data sufficiently wide and diverse to bridge the domain gap between synthetic and real images. Therefore, we can further control the $P(a)$ and $P(\sigma)$ with I configuration parameters $\xi_i\in \Xi$ that the optimization can be parameterized as: 
\begin{equation}
	\begin{aligned}
		& \hat{\theta}=\mathop{\arg\min}\limits_{\theta}\mathbb{E}_{\xi\sim \Xi}\mathbb{E}_{a\sim P_{\xi}(a),\sigma\sim P_{\xi}(\sigma)}\\
		& L[f(\bm{A}_{input}(a_j,\sigma_k^{in});\theta_\Omega)-\bm{A}_{label}(a_j,\sigma_k^{out})].
	\end{aligned}
\end{equation}
in which, $\xi_i$ is bounded as $\xi_i=[\xi_i^{low},\xi_i^{high}]$ and randomly sampled within the range. Hence, the main goal is to model the imaging and determine a reasonable range of configuration parameters to create the parameterized data distribution of parametric templates, $P_{\xi}(a)$, and non-ideal factors, $P_{\xi}(\sigma)$.

A schematic of the MOST-DL framework is shown in Fig. 1. Briefly, we first synthesize the parametric templates, including M$_0$ and T$_2$, from multi-contrast images of a public database (Section \uppercase\expandafter{\romannumeral3}-B). Meanwhile, non-ideal factors are constructed based on physical priors (Section \uppercase\expandafter{\romannumeral3}-C). Depending on the specific task requirements, the model-based simulation will generate the input and corresponding label data with the specific MRI sequences (Section \uppercase\expandafter{\romannumeral3}-D). During data generation, domain randomization is performed to make the synthetic domain sufficiently wide and make the model trained on synthetic data robust enough for real data. As such, the framework can generate paired datasets for various supervised learning tasks, such as under-sampling MRI reconstruction, motion artifact removal and qMRI.

\subsection{Parametric Templates Synthesis}
The quantitative tissue parametric templates were synthesized from the realistic qualitative multi-contrast MR images by the MR signal model:
\begin{equation}
	S=M_0\cdot (1-e^{-TR/T_1})\cdot e^{-TE/T_2}
\end{equation}
Specifically, the PD-weighted image was first assigned as a ‘virtual’ M$_0$ map after intensity normalization under TE → 0 and TR $\gg$ T$_1$. To obtain the other parametric maps, the weighted images are used as $S$ with the corresponding TE/TR value. The parameters distribution can be changed by adjusting the TE/TR value or intensity scaling.

In this work, the multi-contrast images used to produce parametric templates were from the public database IXI \footnote{https://brain-development.org/ixi-dataset/}. It consists of five contrasts collected at three different hospitals in London. For IXI data, the matrix size is 256×256, and the imaging resolution is 0.94 mm × 0.94 mm × 1.25 mm. We randomly selected 200 subjects from Hammersmith Hospital and Guy’s Hospital. The T$_2$-weighted volumes were chosen as references for co-registration by elastix toolbox \cite{belastix} based on Insight Segmentation and Registration Toolkit (ITK) with parameters “translation” and “affine”. Then, two-dimensional (2D) slices were sampled from the registered multi-contrast volumes and performed signal model inversion. For high-accuracy simulation, the parametric templates were interpolated to a matrix size of 512×512 grids. Only slices covering the brain and cerebellum were considered, and about 30 slices were extracted from each subject. Finally, about 6000 slices were used for further model-based simulation. 

\begin{figure*}[h]
	\centering
	\includegraphics[scale=.153]{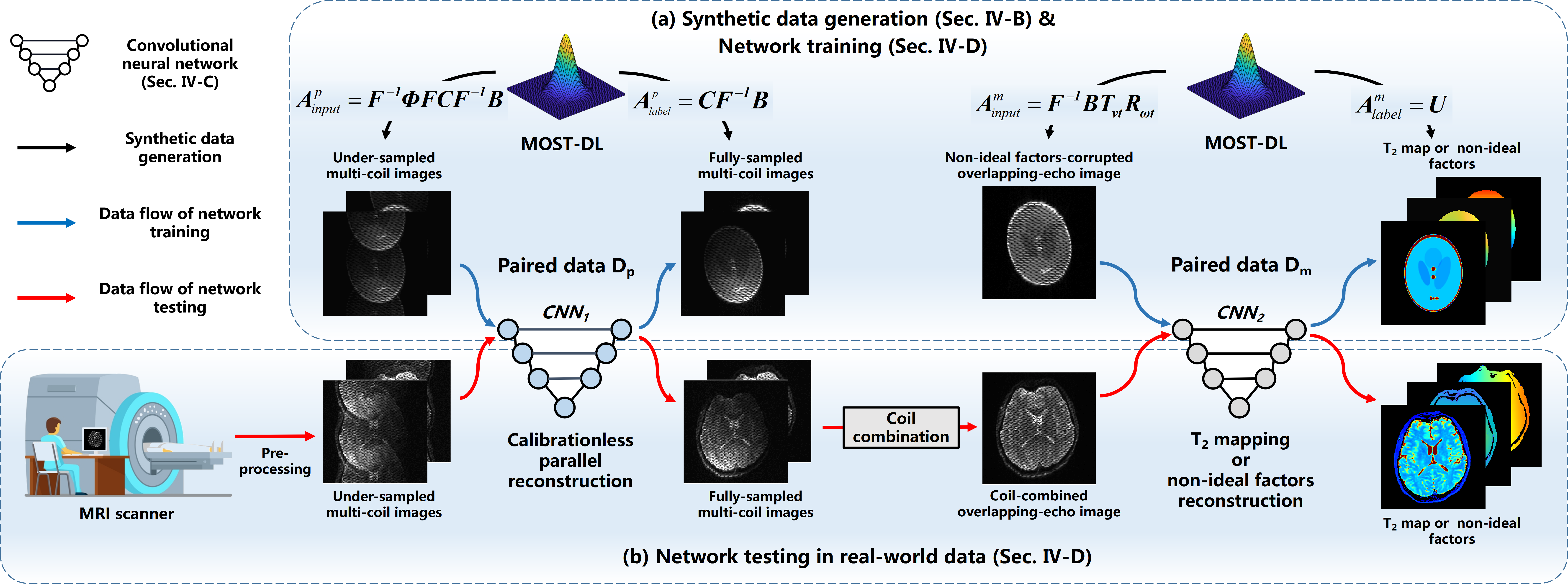}
	\caption{Overview of the proposed pipeline for application in T$_2$ mapping under head motion. (a) Synthetic data generation and network training: paired dataset $D_{p}$ and $D_{m}$ were generated by the MOST-DL framework and used for CNN$_1$ and CNN$_2$ training; (b) Network testing in real-world data: raw data from MRI scanner can be parallel reconstructed and T$_2$ mapping by the trained CNN$_1$ and CNN$_2$. The non-ideal factors (e.g., velocity fields and B$_1^+$ field) can also be reconstructed for visualization. The multi-coil MR images are coil-combined to a single-coil image after CNN$_1$ reconstruction. $F$: Fourier operator; $B$: Bloch equation operator; $U$: down-sampling operation; $T_{vt}R_{\omega t}$: motion operator; $\Phi$: sampling pattern for parallel imaging; $C$: coil sensitivity maps. }
	\label{FIG2}
\end{figure*}

\subsection{Non-ideal Factors Modeling}
\subsubsection{Motion}
The sub-voxel motion of each voxel under Bloch simulation is considered in this framework. A coordinate vector $s_0=[x_0, y_0]$ of parametric templates is created and used to record the accurate position of echo spin (corresponding to the element of the template matrix) at different moments during motion. The additional phase accumulation for each spin at arbitrary evolution time $t$ can be represented as an integral of additional precession frequency caused by motion:
\begin{equation}
		\begin{aligned}
			\Delta \varphi(x,y,t) 
			&=\gamma\int\nolimits_{0}^{t}[G_{RO}(\hat{t})\cdot (x_{\hat{t}}-x)+G_{PE}(\hat{t})\cdot (y_{\hat{t}}-y)]d\hat{t}
		\end{aligned}
\end{equation}
where $\gamma$ is the gyromagnetic ratio, $G_{RO}(t)$ and $G_{PE}(t)$ are the time-varying linear gradient field along the frequency and phase encoding directions, respectively. Hence, all spins with additional phases are finally integrated and contribute to the variation of the acquired signal. Under the assumption of uniform rigid motion during the sequence execution in a single shot, the motion operator $\bm{T_{vt}R_{\omega t}}$ represented by velocities $v_{RO}$, $v_{PE}$ and angular velocity $\omega$ is applied to $s_0$ of each spin to update the coordinate:
\begin{equation}
		\begin{bmatrix}
			x_t \\ y_t
		\end{bmatrix}
	=\bm{T_{vt}R_{\omega t}}
	\begin{bmatrix}
		x_0 \\ y_0
	\end{bmatrix}
\end{equation}
where $\bm{T_{vt}}$ is translation operator, and $\bm{R_{\omega t}}$ is rotation operator at time $t$. The rigid motion parameters can be visualized as velocity fields at pixel level as:
\begin{equation}
	\begin{aligned}
		\begin{array}{ll}
			V_{RO}(x,y)=-\omega\cdot y+v_{RO}\\
			V_{PE}(x,y)=\omega\cdot x+v_{PE}\\
		\end{array}
	\end{aligned}
\end{equation}

\subsubsection{B$_1^+$ inhomogeneity}
The B$_1^+$ (radio frequency field) inhomogeneity is taken as the sum of simple low-order polynomial functions with the random number set $r_p$ and Gaussian functions with the random number set $r_n$:
\begin{small}
	\begin{equation}
		\begin{aligned}
			\Delta B(x,y) &=\sum\limits_{n_x=1}^{N_p}\sum\limits_{n_y=1}^{N_p}r_p(n_x,n_y)x^{n_x}y^{n_y}+\sum\limits_{n_g=1}^{N_g}G(x,y;r_n(n_g))
		\end{aligned}
	\end{equation}
\end{small}
where, $n_x$ and $n_y$ are the order of $x$ and $y$, respectively. $n_g$ represents the superposition of Gaussian profiles. In this work, $N_p$ is set to 2, and $N_g$ is set to 1. Subsequently, $\Delta B$ will be normalized within a reasonable boundary to obtain the final B$_1^+$. The actual flip angle for each spin is calculated as a proportion of the desired flip angle.

\subsubsection{Other non-ideal factors}
The undesirable effects arising from eddy currents, system delays, nonlinear gradient amplifier response function, or even mechanical vibrations can cause gradient imperfections, further resulting in the deviation of acquired k-space from its desired design. We model the gradient imperfections by simulating the random fluctuation of gradient area to cover the comprehensive effect caused by instrument imperfection. Besides, it is common practice to assume that the noise in MRI raw data has a Gaussian distribution with zero mean \cite{bnoise}. Due to the linear and orthogonal nature of the Fourier transform, the real and imaginary images reconstructed from raw data will preserve the Gaussian characteristics of the noise. Therefore, the noise of Gaussian distribution with the same variance is added to the real/imaginary part of the synthetic image. It is possible to expand the framework for other non-ideal factors (e.g., $\Delta$B$_0$, B$_1^-$, chemical shift), and this is something that we are planning for future work.

\subsection{Model-based Simulation and Signal Reconstruction}
The model-based simulation in this framework is based on solving the Bloch equation with a task-specific pulse sequence. By introducing the coil sensitivity maps, the simulation can be extended from a single-coil scenario to a multi-coil scenario. The paired measurements $b_s$ and $b_t$ of different evolution pathways derived from the same tissue relaxation parameters $a_j$ can be obtained by controlling the non-ideal factors $\sigma_k$ and adjusting the forward model $\bm{A}$.

Under the MOST-DL framework, a faithful signal reconstruction relies on physical feasibility, adequate signal representation and decoding ability. To verify the accuracy of data modeling, the MOST-DL provides the possibility of reconstructing non-ideal factors by solving the optimization problem of Equation (5) only with the label changed to non-ideal factors. As such, the non-ideal factors carried in real-world data can be reconstructed explicitly (or visualized) and used as a quality control indicator for secondary validation of the reliability in data generation.

\section{MOLED T$_2$ Mapping under Rigid Motion}
The MOST-DL is applied to build synthetic datasets for MOLED T$_2$ mapping under rigid motion. In this application, the motion correction is jointly achieved by a cascade framework consisting of two CNNs: CNN$_1$ for calibrationless parallel reconstruction to address the mismatch between under-sampled data and ACS data; CNN$_2$ for end-to-end mapping from motion-corrupted MOLED images to motion-free quantitative T$_2$ maps. Fig. 2 shows the data flow of synthetic data generation, network training and testing. The MOLED acquisition and reconstruction are reviewed in Section \uppercase\expandafter{\romannumeral4}-A.    Paired datasets are generated by the MOST-DL according to the forward models as described in Section \uppercase\expandafter{\romannumeral4}-B. Section \uppercase\expandafter{\romannumeral4}-C describes the network architecture used for this application. Finally, the details of network training with synthetic data and testing with real-world data are provided in Section \uppercase\expandafter{\romannumeral4}-D. 
\begin{figure}[!t]
	\centerline{\includegraphics[width=\columnwidth]{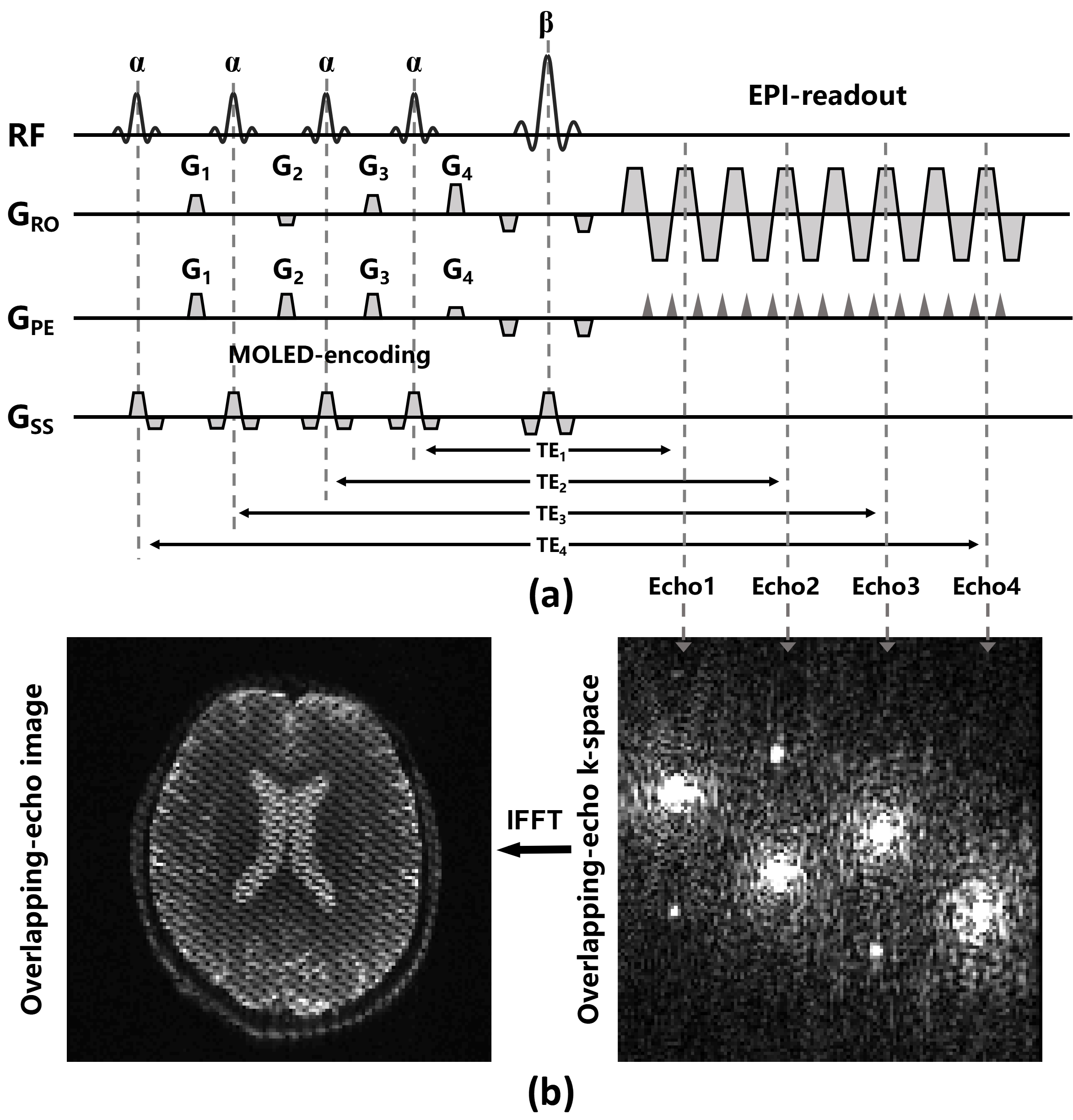}}
	\caption{(a) Single-shot SE-MOLED sequence for T$_2$ mapping. The four TEs of the SE-MOLED sequence are 22.0, 52.0. 82.0, 110.0 ms in this work, corresponding to the four excitation pulses. (b) The overlapping-echo image and k-space.}
	\label{FIG3}
\end{figure}

\subsection{MOLED Acquisition and Reconstruction}
Details of the topic have been presented previously \cite{boled1,boled2}, but a brief summary is provided here. In overlapping-echo acquisition, multiple echo signals containing different information (e.g., relaxation \cite{boled2},  diffusion \cite{boled3}, multi-slices \cite{boledsms}) are encoded in a single k-space to achieve efficient signal compression. These echo signals with different evolution times are prepared by independent RF pulses and are finally acquired with overlapped high-frequency components. The 2D SE-MOLED sequence \cite{boled2}  shown in Fig. 3(a) can be used to acquire echo signals following the T$_2$ signal decay for T$_2$ mapping. Four excitation pulses with the same flip angle $\alpha$ = 30° are followed by a refocusing pulse with a flip angle of $\beta$ = 180° to generate four main T$_2$-weighted spin echoes with different TEs (TE$_1$ = 22.0 ms, TE$_2$ = 52.0 ms, TE$_3$ = 82.0 ms, TE$_4$ = 110.0 ms). The gradients G$_1$, G$_2$, G$_3$ and G$_4$ are echo-shifting gradients used to shift the four echoes away from the k-space center along the phase-encoding and frequency-encoding directions. As shown in Fig. 3(b), the four echo signals with different evolution times are obtained in the same k-space, resulting in an image modulated by interference fringes. We believe that the quantitative information is modulated by these fringes, thus a deep neural network was used to perform direct end-to-end mapping reconstruction without echo separation.

\subsection{Synthetic Data Generation by MOST-DL}
Only T$_2$ and M$_0$ templates were used in synthetic data generation, in which T$_2$$\in$[0, 650] ms, M$_0$$\in$[0, 1]. The T$_1$ value was fixed to 2000 ms for all tissues due to the short duration between the four excitation pulses (about 44 ms). Random rotations (0°, 90°, 180°, 270°) and flips (horizontal and vertical) were applied to the parametric templates for data augmentation. 

Fig. 2(a) shows the pipeline of synthetic data generation relied on the MOST-DL framework. For parallel reconstruction in CNN$_1$, the paired dataset D$_p$ was generated following the forward models $\bm{A}_{input}^p$ and $\bm{A}_{label}^p$ as:
\begin{equation}
	\begin{aligned}
		\left \{
		\begin{array}{ll}
			\bm{A}_{input}^p=\bm{F^{-1}\Phi FCF^{-1}B}\\
			\bm{A}_{label}^p=\bm{CF^{-1}B}\\
		\end{array}
		\right.
	\end{aligned}
\end{equation}
in which, $\bm{F}$ is the Fourier operator, $\bm{B}$ is the Bloch operator for 2D SE-MOLED sequence, $\bm{\Phi}$ is the sampling pattern, $\bm{C}$ is the coil sensitivity maps. Due to the nature of EPI readout, a uniform under-sampling with the central region not fully-sampled was used as $\bm{\Phi}$, and the acceleration rate R = 2. The multi-coil images were generated from the multiplication of synthetic single-coil overlapping-echo images and coil sensitivity maps obtained from offline collected ACS data. These ACS data were collected by conventional GRAPPA scanning protocol. A sensitivity maps pool containing about 100 slices was generated from scans of 5 healthy volunteers using the ESPIRiT algorithm \cite{bespirit}. For end-to-end T$_2$ mapping and non-ideal factors reconstruction in CNN$_2$, the paired dataset D$_m$ was generated following the forward models $\bm{A}_{input}^m$ and $\bm{A}_{label}^m$ as:
\begin{equation}
	\begin{aligned}
		\left \{
		\begin{array}{ll}
			\bm{A}_{input}^m=\bm{F^{-1}BT_{vt}R_{\omega t}}\\
			\bm{A}_{label}^m=\bm{U}\\
		\end{array}
	\right.
	\end{aligned}
\end{equation}
where $\bm{U}$ is the down-sampling operation (applied on spin-level parametric templates or non-ideal factors). As mentioned above, the rigid motion as a main non-ideal factor can be described by the motion operator $\bm{T_{vt}R_{\omega t}}$. The corresponding T$_2$ templates, velocity fields and B$_1^+$ with size of 512×512 were down-sampled to 256×256 as labels.  During Bloch simulation, all RF pulses were simulated using hard pulses with spatial B$_1^+$ inhomogeneity. Gradient fluctuation was applied in MOLED echo-shifting gradients. The step size in time was 0.003 ms for readout gradients and 0.1 ms for other gradients. GRAPPA was not considered in synthetic data, and the echo spacing (ESP) of readout gradients was 1/R of that in the \emph{in vivo} experiment to maintain a consistent echo train length (ETL). The detailed imaging parameters were ESP = 0.465 ms, field of view (FOV) = 22 × 22 cm$^2$, and matrix size = 128 × 128. Gaussian noise was added in single-/multi-coil overlapping-echo images.

\begin{figure}[!t]
	\centerline{\includegraphics[width=\columnwidth]{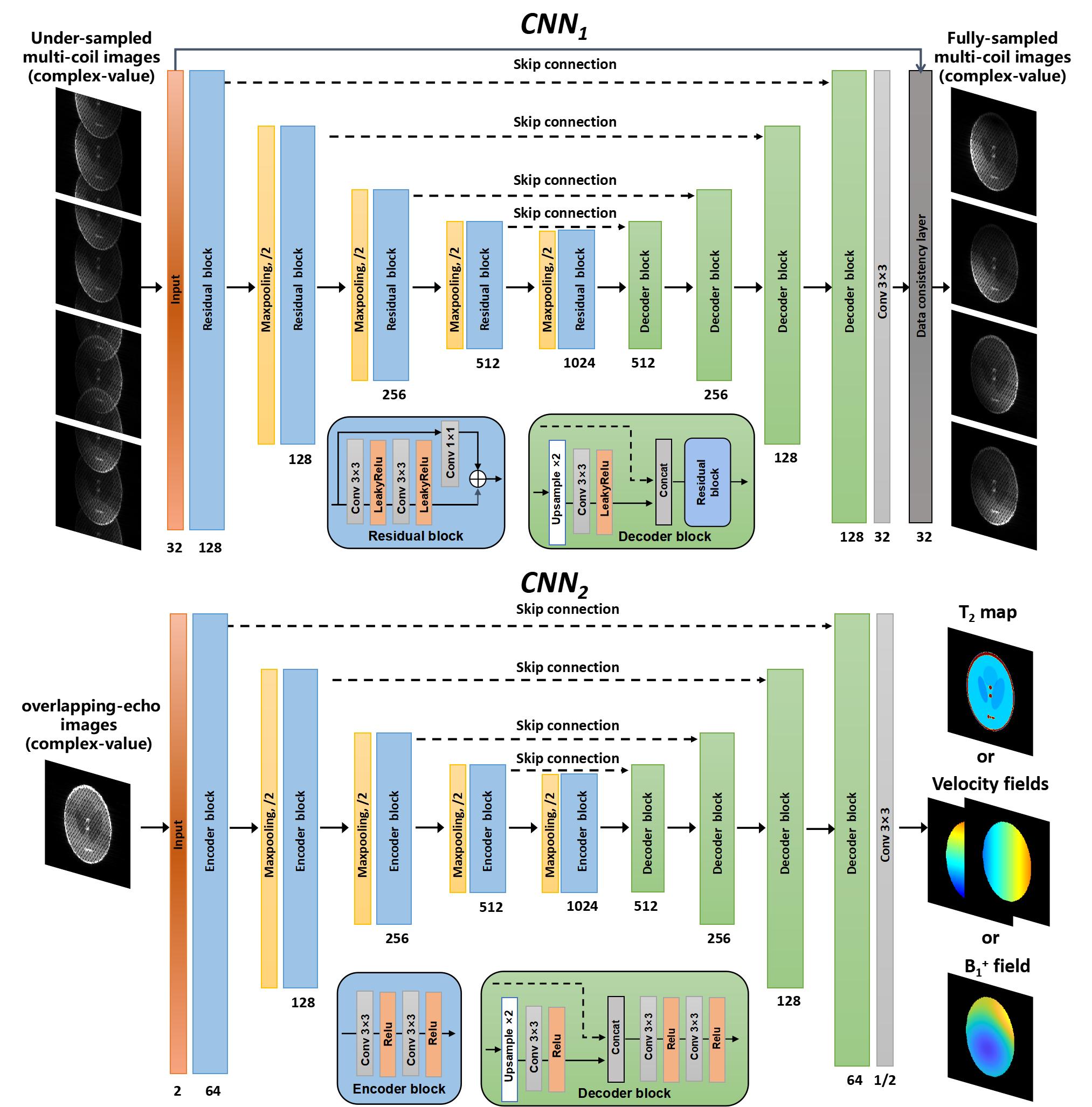}}
	\caption{The proposed CNN$_1$ and CNN$_2$ architectures. The network backbone is based on U-Net, which consists of a series of encoder blocks and decoder blocks.}
	\label{FIG4}
\end{figure}
For domain randomization, we randomized the following aspects of the synthetic domain:
\begin{itemize}
	\item Distribution of T$_2$ value of parametric templates;
	\item SNR of multi-coil/single-coil MR images: 30.0 to $\infty$ dB;
	\item Gradient fluctuation for MOLED echo-shifting gradients: -5\% to 5\%; 
	\item B$_1^+$ inhomogeneity of excitation pulses: 0.7 to 1.2; 
	\item The velocities $v_{RO}$ and $v_{PE}$: -10.0 to 10.0 cm/s, and the angular velocity, $\omega$: -50.0 to 50.0 °/s; 
	\item Randomly matching of coil sensitivity maps and synthetic single-coil images for generating multi-coil images;
\end{itemize}
Other factors were considered to have no significant contribution to these two tasks and were therefore ignored. 

Finally, 8,000 paired samples (under-sampled multi-coil images vs. fully-sampled multi-coil images) were generated for CNN$_1$ training, and 15,000 paired samples (overlapping-echo images vs. T$_2$ maps/velocity fields/B$_1^+$ fields) were employed for CNN$_2$ training. The data generation took approximately 25 hours. The Bloch simulation was implemented in MRiLab \cite{bmrilab} and SPROM software \cite{bsprom} on a PC with an NVIDIA GeForce RTX 2080 Ti GPU. Other processes were performed using MATLAB (R2019b) software (Mathworks, Natick, MA, USA). 

\subsection{Network Architecture}
Our network backbone is based on a five-level U-Net \cite{bunet}, which consists of a series of encoder blocks to extract high-dimensional features from original MR images and decoder blocks to reconstruct target signals. The detailed CNN$_1$ and CNN$_2$ architectures are shown in Fig. 4. In CNN$_1$, a residual learning block is used as the encoder block, and a data consistency layer \cite{bdc} is introduced for parallel reconstruction. The value of empirical parameter $\lambda$ of the data consistency layer is set to 1.0 for denoising, which represents the reconstructed result is the combination of the CNN prediction and the original measurement. In both CNN$_1$ and CNN$_2$, up-sampling operations in the decoder blocks were carried out through bilinear interpolation instead of up-convolution. The final output was generated using the last 3×3 convolution layer without activation function. The trainable parameters for CNN$_1$ and CNN$_2$ were 52.7 M and 34.5 M, respectively. 

\subsection{Training and Testing Details}
Fig. 2(a) illustrates the data flow of network training. Parallel reconstruction and end-to-end T$_2$ mapping tasks both affect the final result but are independent of each other, so we trained CNN$_1$ and CNN$_2$ separately using datasets $D_p$ and $D_m$, respectively. For CNN$_2$, the non-ideal factors reconstruction only serves as visual quality control and does not affect T$_2$ mapping. Therefore, the same network structure with different parameters was used to map from overlapping-echo images to different modalities (T$_2$ map, velocity fields or B$_1^+$ field). Besides, before being fed into CNN$_2$, the overlapping-echo image (128×128) was first zero-padded in k-space to 256×256 and then normalized by the maximum value of magnitude in the image domain. The paired samples were randomly cropped into 96×96 patches during the CNN$_2$ training phase because the MOLED echo signals with different evolution times were encoded in the local modulation. However, the patching operation is not necessary for the testing phase due to the sliding window manner of convolution. 

For both CNN$_1$ and CNN$_2$, the paired synthetic data sets were randomly split into 90\% and 10\% for training and validation. The complex-valued multi-/single-coil overlapping-echo images were divided into real and imaginary components as two individual channels for the network input \cite{bdc}. We used $l_1$ norm as the loss function and Adam optimizer with momentum parameters $\beta_1$ = 0.9 and $\beta_2$ = 0.999 to update network parameters. The initial learning rate was 10$^{-4}$, which decreased by 20\% after each 80,000 iterations until the network converged. The network training for CNN$_1$ and CNN$_2$ took about 12 and 20 hours, respectively. Finally, the best models of CNN$_1$ and CNN$_2$ with the lowest loss on the validation set were selected for testing purposes.

The data flow of network testing is shown in Fig. 2(b). The raw data acquired from the MRI scanner was first pre-processed, including intensity scaling and 3-line linear phase correction to remove EPI Nyquist ghosting. The multi-coil data reconstructed from network CNN$_1$ were coil-combined by an adaptive coil combination algorithm \cite{badp}, in which the coil with the highest SNR was selected as the reference coil. Before being fed into CNN$_2$, the coil-combined 128×128 overlapping-echo image was also zero-padded to 256×256 in k-space and then normalized in the image domain. The network training and testing were implemented in Python using the PyTorch library on a PC with an NVIDIA GeForce RTX 2080 Ti GPU. The pre-processing and coil combination for real-world data were performed using MATLAB (R2019a) software (Mathworks, Natick, MA, USA).

\begin{figure*}[h]
	\centering
	\includegraphics[scale=.168]{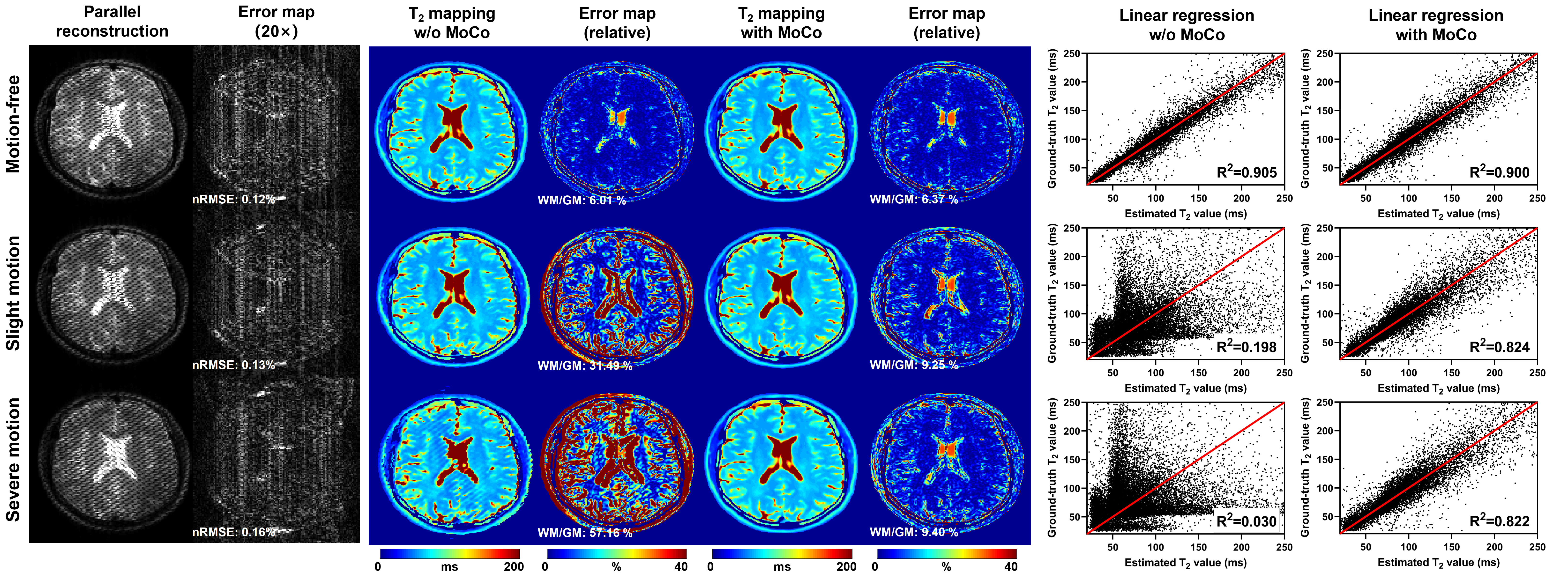}
	\caption{Parallel reconstruction and T$_2$ mapping results of numerical brain using the MOST-DL method from data with motion-free (row 1), slight motion (row 2) and severe motion (row 3). The T$_2$ range of linear regression analysis is 20 to 250 ms. Slight motion: $v_{RO}$ = -2.0 cm/s, $v_{PE}$ = -2.0 cm/s, $\omega$ = -10.0 °/s; Severe motion: $v_{RO}$ = -8.0 cm/s, $v_{PE}$ = -5.0 cm/s, $\omega$ = -32.0 °/s. MoCo: Motion correction.}
	\label{FIG5}
\end{figure*}
\subsection{Validation Experiments}
The study protocol was approved by the institutional research ethics committees, and written informed consents were obtained from the volunteers and the patient’s guardians prior to the experiments.

\subsubsection{Numerical Human Brain Experiments}
We first conducted numerical human brain experiments with known quantitative parameters. The original parametric templates were also generated from a multi-contrast volume selected from the IXI database following the MOST-DL pipeline. The parametric templates, including T$_2\in$[0, 600] ms and M$_0\in$[0, 1], were used as the ground truth to evaluate the reconstruction performance. The forward operators in Equations (11) and (12) were applied to obtain single/multi-coil overlapping images for network testing. The imaging parameters were consistent with that of training data, and Gaussian noise was added in the numerical brain to achieve the SNR of 34.0 dB.

\subsubsection{In Vivo Experiments}
The \emph{in vivo} experiments in this section were conducted on a whole-body MRI system at 3T (MAGNETOM Prisma TIM, Siemens Healthcare, Erlangen, Germany) with a 16-channel head coil. All motion-related \emph{in vivo} data were acquired from four healthy volunteers and a patient with epilepsy using the SE-MOLED sequence. The healthy volunteers were instructed for three scans: (1) reference scan, (2) motion-free scan and (3) continuous motion scan. The reference scan was employed only once at the beginning of the whole scan time to obtain ACS data. The (2) and (3) scans used parallel imaging and the acceleration factor R = 2. In the motion-corrupted scan, the subjects were asked to randomly move their head. This scan was repeated several times, with each session lasting 80 s. Besides, a healthy volunteer was instructed for an additional scan with continuous nodding to explore the performance of the proposed method under through-plane motion. The patient data were obtained by appending the SE-MOLED sequence in a standard clinical exam. The relevant imaging parameters include FOV = 22×22 cm$^2$, matrix size = 128×128, slice thickness = 4 mm, slice number = 21, ESP = 0.93 ms. For comparison, a conventional spin-echo (SE) sequence was acquired on the four healthy volunteers with parameters: TEs = 35, 50, 70, 90 ms. We also collected additional motion-free SE-MOLED data from another 15 healthy volunteers for network training in comparison methods.

\subsubsection{Comparative Algorithms}
We used two calibration-based parallel reconstruction methods (GRAPPA \cite{bepigrappa} and ESPIRiT \cite{bespirit}) and three calibrationless methods (SAKE \cite{bepi2}, ALOHA \cite{bepi3}, and real data-driven deep learning \cite{bepi5}) to verify the performance of our parallel reconstruction method in \emph{in vivo} experiments. Due to the difficulty in obtaining the fully-sampled ground truth of the SE-MOLED sequence, we used the GRAPPA reconstructed results as labels for the real data-driven deep learning method, and the CNN$_1$ was trained for a fair comparison. For motion correction, we conducted comparative experiments using different motion simulation strategies. Image-domain simulation strategy (similar to Johnson \emph{et al.} \cite{bmotionimage}) and k-space simulation strategy (similar to Duffy \emph{et al.} \cite{bmotionksp}) were used as comparative methods. The CNN$_2$ was selected as the motion correction network for all simulation strategies. The simulation parameters of velocities $v_{RO}$, $v_{PE}$ and angular velocity $\omega$ were consistent with that of MOST-DL. Additionally, self-comparison experiments of domain randomization were conducted to evaluate the impact of noise, B$_1^+$ inhomogeneity, gradient fluctuation, T$_2$ distribution and motion correction. 

\section{Results}
\subsection{Experiments with Numerical Human Brain}
In Fig. 5, the results of parallel reconstruction (CNN$_1$) and T$_2$ mapping (CNN$_2$) under different levels of rigid motion are plotted. In all cases, the parallel reconstruction results show high quality with normalized root mean square error (nRMSE) values below 0.2\%. For the motion-free case, the final T$_2$ maps reconstructed with/without motion correction are similar in both the quantitative maps and error maps. With the inclusion of motion, the T$_2$ maps without motion correction become corrupted, causing a higher error compared with the ground truth. In contrast, the motion-corrected maps remain high quality with low error levels (\textless10\% relative error) in gray/white matter (GM/WM). These results are supported by linear regression analysis. The R$^2$ values show significant improvement after motion correction (from 0.198 to 0.824 in slight motion case, from 0.030 to 0.822 in severe motion case).

\begin{figure*}[h]
	\centering
	\includegraphics[scale=.205]{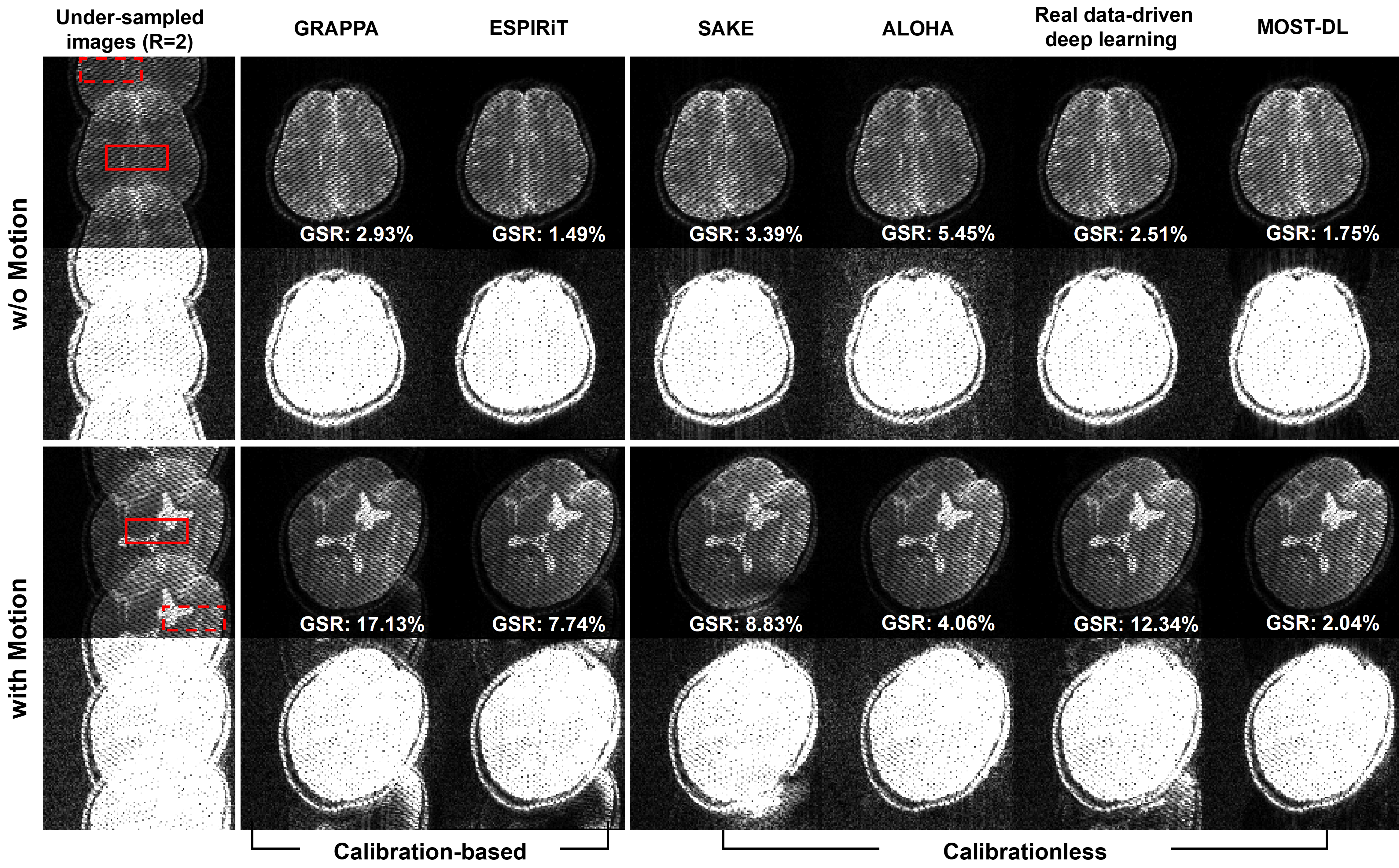}
	\caption{Parallel reconstruction results of under-sampled SE-MOLED images by various methods. The ten times re-scaled images are shown below the original images. The GSR values were calculated using the mean magnitude in regions marked by red solid boxes (signal) and red dotted boxes (ghost).}
	\label{FIG6}
\end{figure*}
\subsection{Experiments with Real Data}
Fig. 6 shows the parallel reconstruction results of \emph{in vivo} human brain using various comparison methods and the proposed MOST-DL-based method (with CNN$_1$). To compare the results quantitatively, we also calculate the ghost-to-signal ratio (GSR) value. For the motion-free case, both calibration-based and calibrationless methods performed well and had low GSR values. However, in the motion-corrupted cases, significant artifacts appear in calibration-based results due to the mismatch between the reference scan and motion-corrupted scan. SAKE and real data-driven deep learning methods also face challenges in motion-corrupted cases that visible artifacts are presented in scaled images. Both the proposed method and ALOHA eliminated all visible artifacts, however, compared with MOST-DL, ALOHA has a higher GSR value in motion and motion-free cases.

\begin{figure*}[h]
	\centering
	\includegraphics[scale=.115]{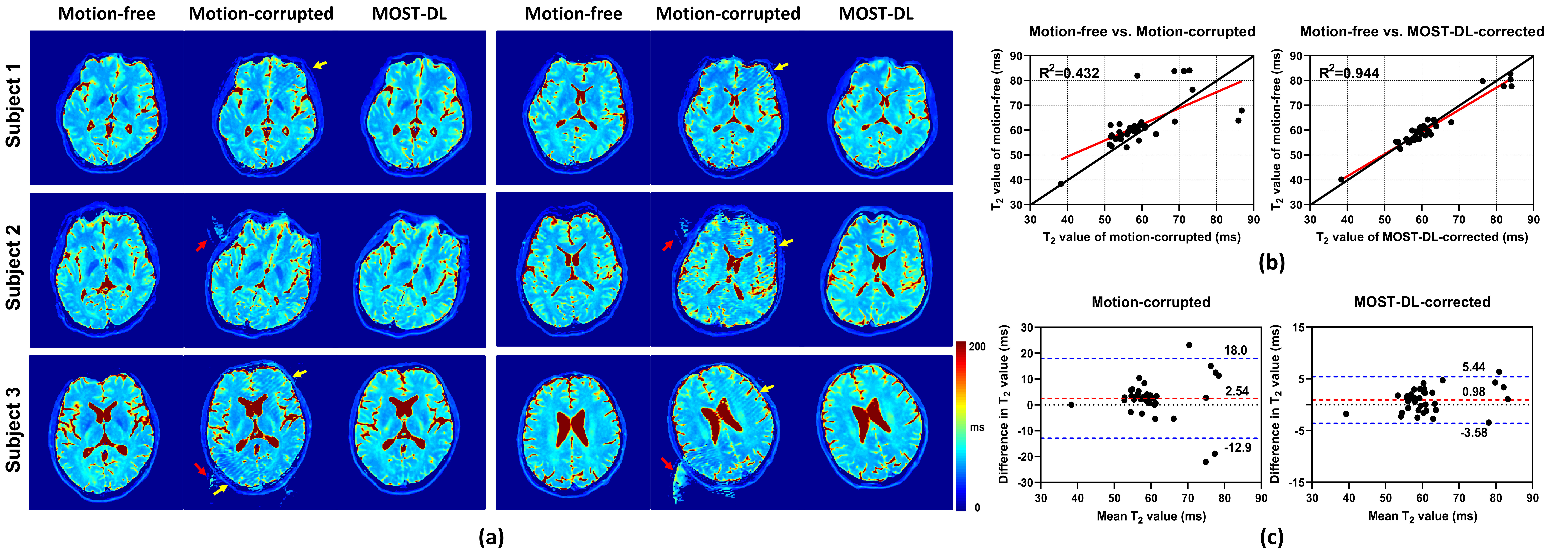}
	\caption{T$_2$ mapping results of \emph{in vivo} data. (a) T$_2$ mapping results of different slices with in-plane motion from three subjects. The ghosting artifacts are marked by red arrows and the motion artifacts are marked by yellow arrows. (b) Linear regression plots comparing T$_2$ derived from motion-free cases vs. motion-corrupted cases and MOST-DL-corrected cases. (c) Bland-Altman plots corresponding to the linear regression plots shown in (b). In the Bland-Altman plots, the blue dotted lines represent 95\% confidence level, and the red dotted lines represent mean T$_2$ value differences.}
	\label{FIG7}
\end{figure*}
Fig. 7(a) illustrates the results of T$_2$ mapping (with CNN$_2$) from 3 healthy volunteers. One can see that the motion-corrupted cases (parallel reconstruction by GRAPPA) suffered from ghosting artifacts (red arrows) and motion artifacts (yellow arrows). With the application of the proposed MOST-DL to parallel reconstruction and motion correction, these artifacts are eliminated, and the image quality is significantly improved compared with motion-corrupted cases. Quantitative analysis of T$_2$ values between motion-corrupted/motion-corrected cases and motion-free cases is shown in Fig. 7(b, c). The T$_2$ values were calculated from 36 regions of interest (ROIs, 12 ROIs of each subject) placed within the regions most affected by motion artifacts (globus pallidus, frontal white matter and insular cortex) after registration. The regression plots in Fig. 7(b) show better consistency between motion-corrected cases and motion-free cases (R$^2$ = 0.944) compared with motion-corrupted cases (R$^2$ = 0.432). The Bland-Altman plots (Fig. 7(c)) support these results that the motion-corrupted cases show a broader range of differences: motion-corrupted cases: mean difference = 2.54 ms, upper and lower limits of agreement = -12.9 ms and 18.0 ms; motion-corrected cases: mean difference = 0.98 ms, upper and lower limits of agreement = -3.58 ms and 5.44 ms.

\subsection{Effects of Parametric Templates}
The reconstructed T$_2$ maps of human brain from SE-MOLED using different training samples (geometrical shapes vs. public database) are shown in Fig. 8. We can see that the reconstructed texture from a public database is more similar to the reference. Artifact (red arrow) and over-smoothing (red dotted circle) appear in the geometrical shapes cases.

\subsection{Effects of Motion Simulation Strategy}
To verify our claim that high-precision motion simulation plays a key role in motion correction and T$_2$ mapping, we compared our proposed method with various motion simulation strategies. Note that the multi-coil MOLED images have been parallel reconstructed by trained CNN$_1$. As shown in Fig. 9(a-d), signal corruption (yellow arrows) and signal loss (green arrows) appear in retrospective image-domain and k-space motion simulation methods. In contrast, the proposed MOST-DL (Bloch-based simulation) gives the closer results to the real-world data. In Fig. 9(e-h), we can see that there are still residual motion artifacts by using the retrospective motion simulation method (f, g), and the reconstruction even fails in some cases. We believe that the inaccurate motion simulation is the main source of error in the final T$_2$ mapping results.

\begin{figure}[!t]
	\centerline{\includegraphics[width=\columnwidth]{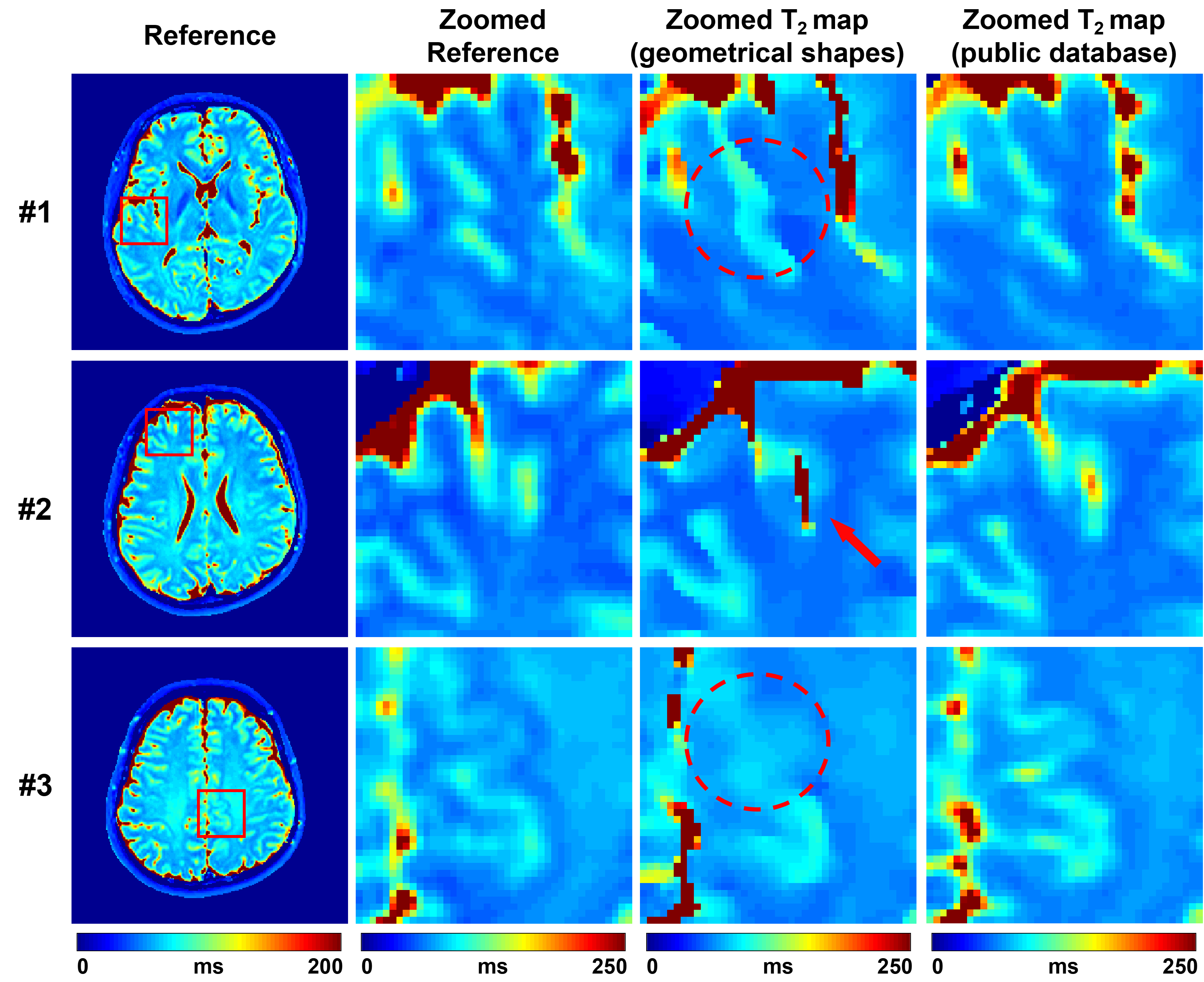}}
	\caption{T$_2$ mapping results of three representative slices reconstructed using different training samples (geometrical shapes vs. public database). In the geometrical shapes cases, the artifact is marked by red arrow, and the over-smoothing is marked by red dotted circle. }
	\label{FIG88}
\end{figure}

\begin{figure}[!t]
	\centerline{\includegraphics[width=\columnwidth]{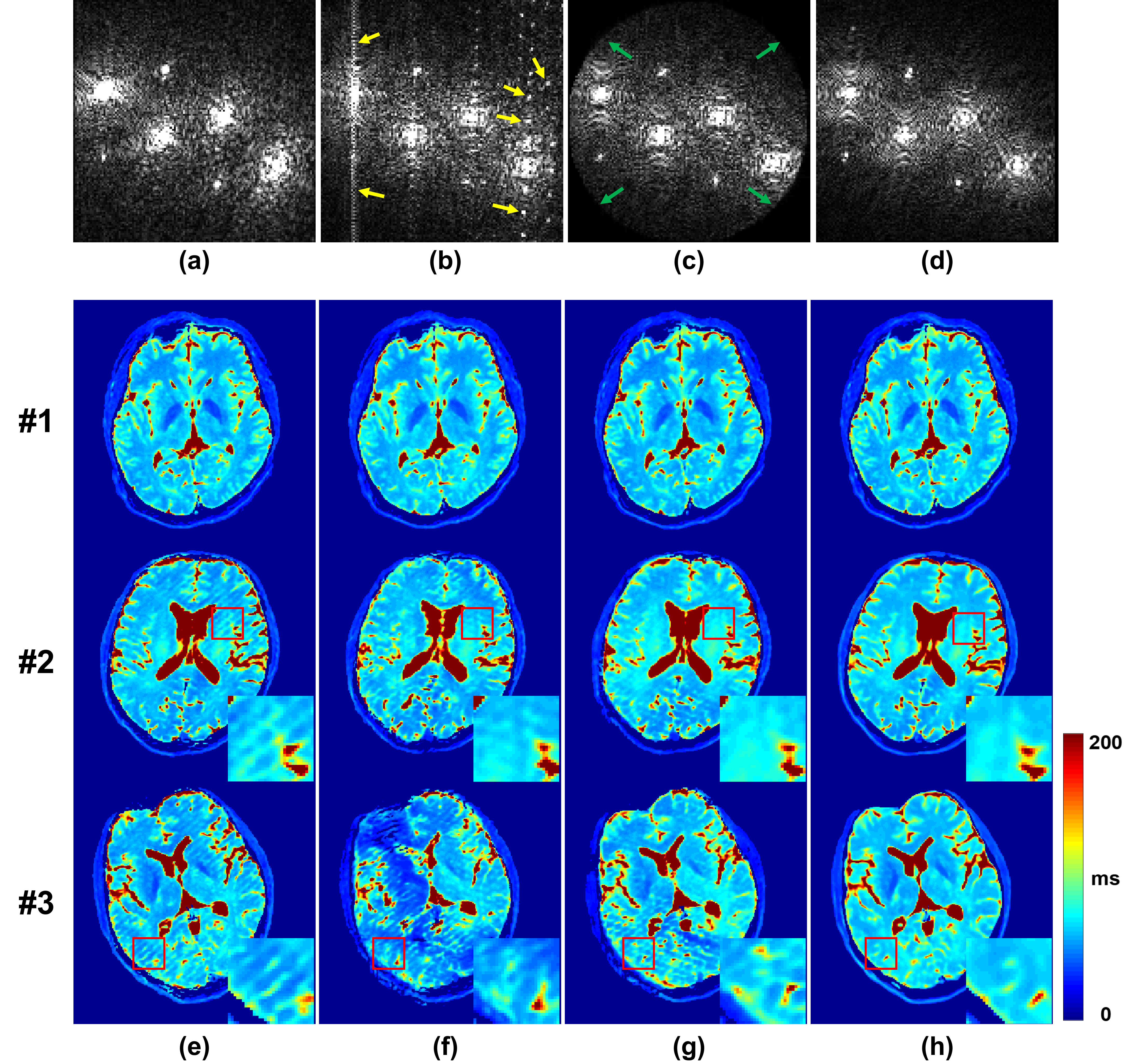}}
	\caption{Overlapping-echo k-space of (a) real-world motion-corrupted case, (b) image-domain motion simulation, (c) k-space motion simulation and (d) Bloch-based motion simulation. The T$_2$ maps reconstructed from CNN$_2$ (e) without motion correction and with (f) image-domain simulation, (g) k-space simulation and (h) MOST-DL training samples in motion-free (upper row) and two motion cases (middle and lower rows).}
	\label{FIG8}
\end{figure}
\subsection{Effects of Domain Randomization}
Here, we verify that the domain randomization during the data generation stage affects the final MOLED T$_2$ mapping results. The reference T$_2$ maps were obtained using the SE sequence. The quantitative analysis (linear regression) is presented in TABLE 1 from manually segmented ROIs (thalamus, caudate nucleus, putamen, globus pallidus, frontal white matter and insular cortex) of 3 healthy volunteers in motion-free results. The full domain randomization exhibits the highest R$^2$ value of linear regression. As for motion-corrupted cases in Fig. 10(a), considerable motion artifacts remain in the T$_2$ maps without motion randomization. These motion artifacts are obliquely striped and primarily distributed in the region of frontal white matter and insular cortex. The mean and variance T$_2$ value curves in Fig. 10(b) shows that the stability of the results without motion randomization is significantly lower and accompanied with greater variance, which means that motion artifacts heavily influence the T$_2$ values within the ROIs.

\begin{table}[h]
	\centering
	\renewcommand\arraystretch{1.1}
	\caption{Self-comparison of domain randomization (DR)}
	\begin{tabular}{cccc}
		\toprule
		\multirow{2}{*}{\textbf{Evaluation type}} & \multicolumn{3}{c}{\textbf{R$^2$ of linear regression}}      \\     \cmidrule{2-4}
		& \textbf{Subject 1} & \textbf{Subject 2} & \textbf{Subject 3} \\ \cmidrule{1-4}
		Full DR                                   & \textbf{0.981}     & \textbf{0.930}     & \textbf{0.988}     \\
		w/o B$_1^+$ inhomogeneity                     & 0.976              & 0.918              & 0.952              \\
		w/o Noise added                           & 0.980              & 0.923              & 0.988              \\
		w/o Gradient fluctuation                        & 0.980              & 0.901              & 0.970              \\
		w/o Random T$_2$ distribution                & 0.969              & 0.911              & 0.975              \\ \bottomrule
	\end{tabular}
\end{table}
\begin{figure*}[h]
	\centering
	\includegraphics[scale=.28]{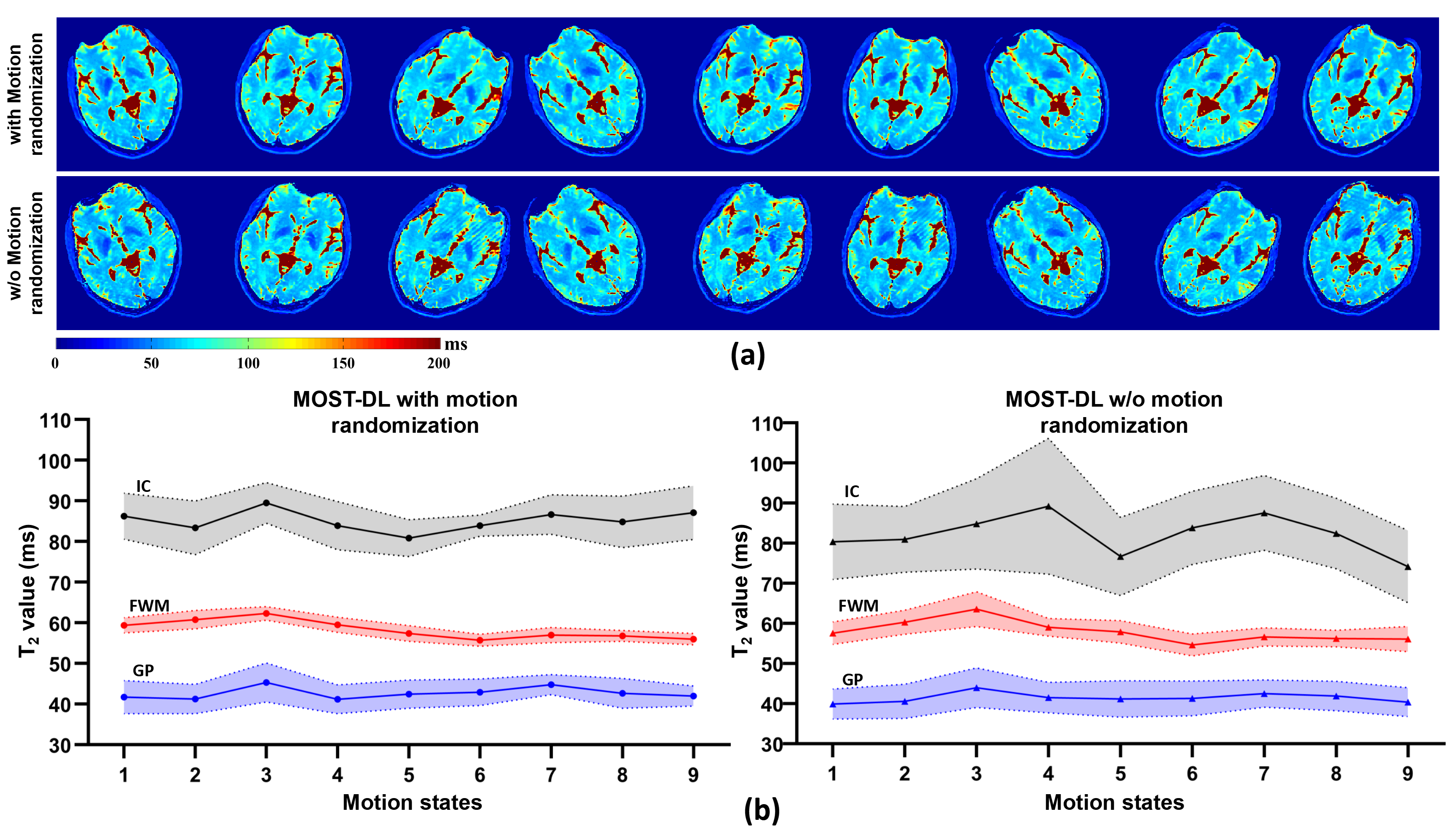}
	\caption{Self-comparison of domain randomization in rigid motion. (a) Sequential T$_2$ maps of representative slice produced by MOST-DL with and without motion randomization. (b) Mean and variance T$_2$ value curves from 3 ROIs of 9 motion states in (a). IC: Insular cortex; FWM: Frontal white matter; GP: Globus pallidus.}
	\label{FIG9}
	\vspace{-3mm}
\end{figure*}
\begin{figure}
	\centerline{\includegraphics[width=\columnwidth]{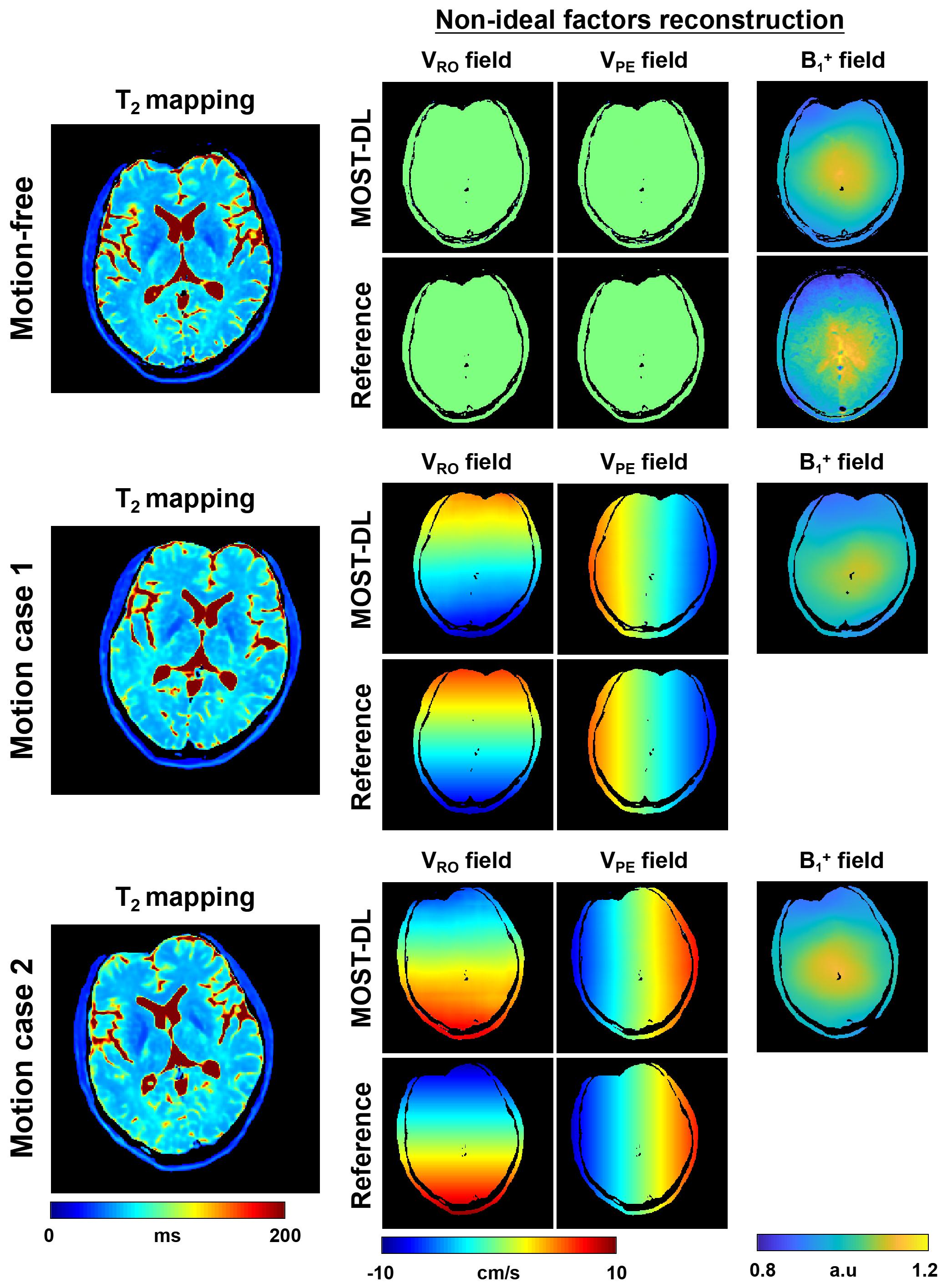}}
	\caption{Non-ideal factors reconstruction results (velocity fields and B$_1^+$ fields left to right) of a representative slice during different states in motion-free (upper row) and two motion cases (middle and lower rows).}
	\label{FIG10}
\end{figure}
\subsection{Reconstruction of Non-ideal Factors}
As secondary validation of the reliability of our method in data generation, the non-ideal factors (velocity fields, B$_1^+$ field) were reconstructed by retrained network CNN$_2$. To obtain references of velocity fields, more of the same echo trains and the refocusing pulses were intentionally appended to the original SE-MOLED sequence, resulting in four MR images to record the subject motion. Subsequently, four MR images are used to calculate three sets of parameters of rigid motion (i.e., translation (mm) along the $x$ and $y$ directions and rotation (degree)) using Statistical Parametric Mapping (SPM) software. Then, $v_{RO}$,$v_{PE}$, and $\omega$ are obtained by regressing the motion parameters and the time between excitation pulses and refocusing pulses. The reference velocity fields are generated according to Equation (9). For B$_1^+$ fields, the references were obtained using the Siemens product B$_1^+$ map based on the turbo-flash sequence. Fig. 11 illustrates the reconstructed velocity fields, B$_1^+$ field and the corresponding references from a same slice during different motion states. We can see that both the predicted results agree well with their references.

\subsection{Effects of Through-plane Motion}
Although through-plane motion correction is challenging for 2D pulse sequences, we also explored its effect of it on the current method. To capture the through-plane motion synchronously with T$_2$ mapping, the SE-MOLED sequence with four echo trains was also used as mentioned above. The through-plane velocity was estimated based on the duration of each echo train and the change in signal strength relative to the motion-free case. When the signal is abnormally attenuated (or disappeared), we assume that through-plane motion beyond the slice thickness has occurred during the time interval between the excitation pulse and refocusing pulse. The excitation slice thickness is 4.0 mm, while the refocusing slice thickness is 3.0 mm. All assessments were performed under the assumption that the subject was nodding at a uniform velocity due to the narrow sampling window within 300 ms. Fig. 12 illustrates the results of T$_2$ mapping under such through-plane motion. The T$_2$ maps were reconstructed with good image quality under slight (\textless1.5 cm/s) and medium (1.5 cm/s$\sim$3.5 cm/s) through-plane motion. Severe through-plane motion (\textgreater 3.5 cm/s) strongly impacts original MRI signals and degrades the final T$_2$ map.

\begin{figure}
	\centerline{\includegraphics[width=\columnwidth]{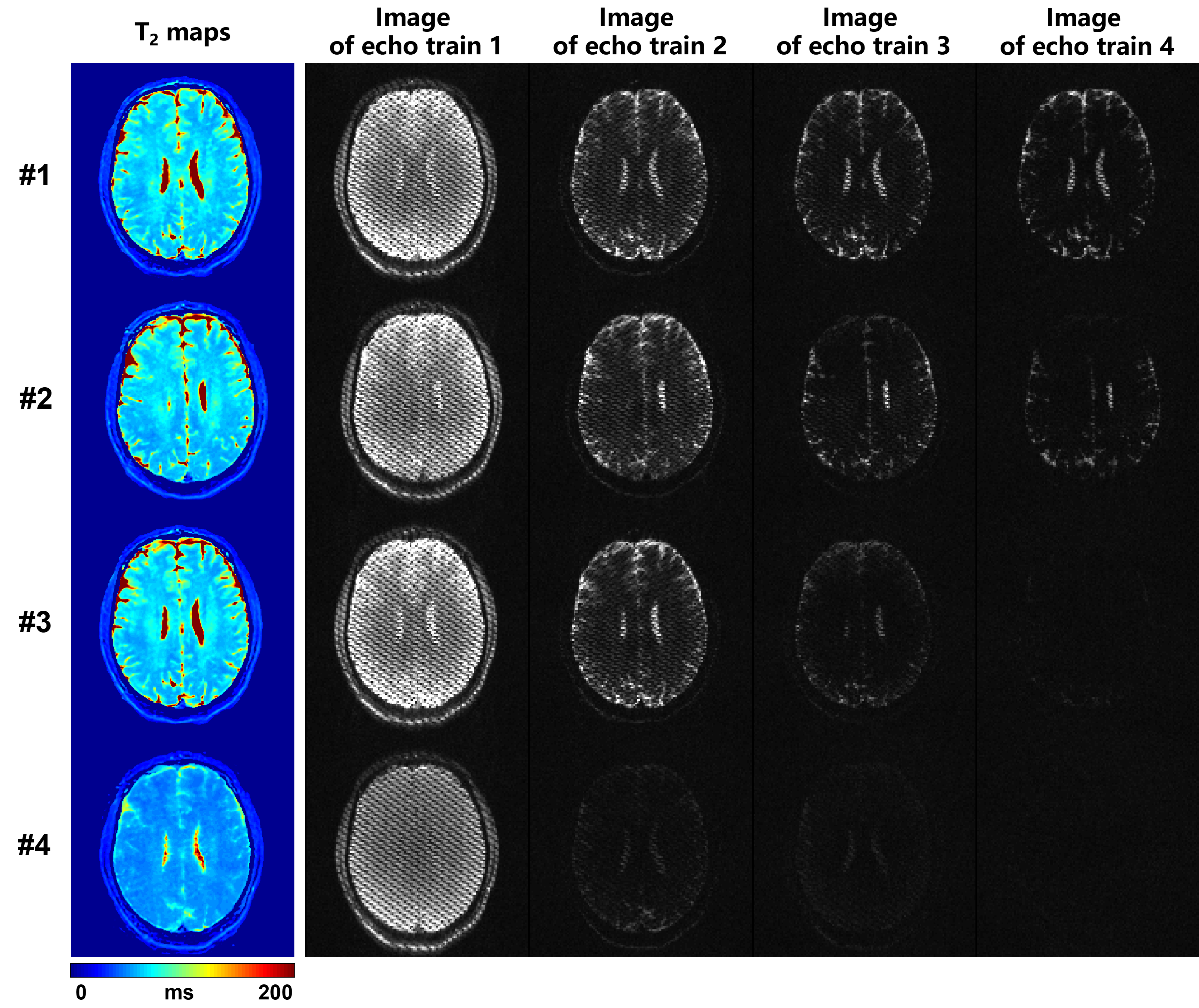}}
	\caption{T$_2$ mapping and the corresponding overlapping-echo images of four echo trains under through-plane motion. From top to bottom: cases of motion-free, slight motion: \textless1.5 cm/s, medium motion: 1.5 cm/s$\sim$3.5 cm/s, and severe motion: \textgreater3.5 cm/s.}
	\label{FIG11}
\end{figure}

\subsection{An Example of A Clinical Case}
Fig. 13 shows the results of a 10-year-old patient with epilepsy. Strong streak artifacts from motion are observed in the T1WI (MPRAGE sequence, Fig. 13(a)) and T2WI (TSE sequence, Fig. 13(b)), which present challenges to quantitative measurement of hippocampal T$_2$. The T$_2$ mapping results from the SE-MOLED sequence are shown in Fig. 13(c, d). We can see that the motion-corrupted case suffered from severe ghosting artifacts and the MOST-DL-corrected case achieved T$_2$ maps with high quality without artifacts. Since the patient's motion occurred randomly, it is difficult to evaluate the motion level during the whole scanning time. However, we observe that the SE-MOLED combined with MOST-DL is more robust to unpredictable motion compared with the multi-shot acquisition.

\begin{figure}
	\centerline{\includegraphics[width=\columnwidth]{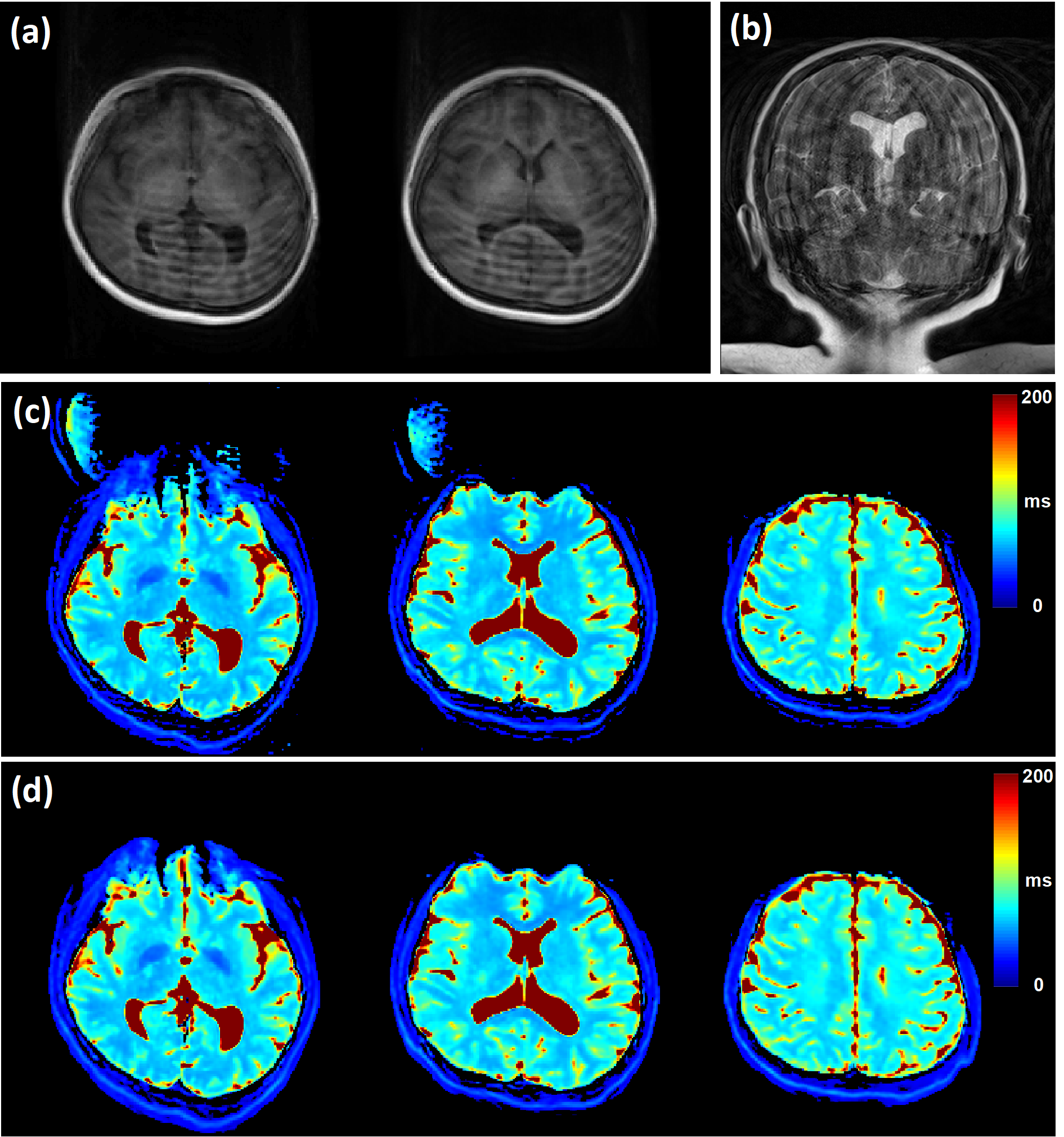}}
	\caption{The results of a 10-year-old patient with epilepsy. (a) MR images from T1 MPRAGE sequence. (b) MR image from T2 TSE sequence. (c) The MOLED T$_2$ maps without motion correction. (d) The MOLED T$_2$ maps reconstructed by the MOST-DL method.}
	\label{FIG12}
\end{figure}

\section{Discussion}
\subsection{Learning from Synthetic Data}
In this work, we developed a synthetic data generation framework to solve a challenging quantitative MRI problem under severe head motion. The neural network is trained with synthetic datasets and can be well generalized to \emph{in vivo} experimental data without network fine-tuning. Two factors are considered to play crucial roles, i.e., (1) generating data using rich anatomical texture priors from a public database, (2) the accurate modeling of the forward operator and non-ideal factors (especially subject motion in the Bloch simulation) with domain randomization. The tissue relaxation parameters in previous studies were created by randomly filling blank templates with hundreds of different basic geometric shapes, which can render the texture of the reconstruction results match the real situation poorly. Moreover, accurate modeling makes the data distribution in the synthetic domain closer to that in the real domain. With domain randomization, discrepancies between the synthetic and real data are modeled as variability, further making the data distribution of the synthetic domain sufficiently wide. Unlike learning from real data, MOST-DL is not limited to acquisition methods and experimental instruments but is built on signal models. This makes the whole framework more generalizable and flexible.

Recently, several deep learning methods have been proposed to focus on reconstruction and motion correction for ultra-fast imaging sequences (e.g., single-shot EPI \cite{bepi5} or multi-shot EPI \cite{bmsepi1}). Due to the difficulty in obtaining fully-sampled or motion-free ground truth, the reconstructed or motion-corrected results with traditional iterative algorithms are usually used as labels for network training. In this work, the proposed MOST-DL method makes it possible to produce perfect data pairs according to the forward model, with the flexibility to increase the diversity of the training data. As shown in Fig. 6, we compared the parallel reconstructed results of the human brain using real data (reconstructed labels) and synthetic data. The network trained from real data shows excellent performance in the motion-free cases but degradation in the motion cases. We believe that the reconstruction errors may caused by imperfect training data pairs and limited data patterns.

In MRI simulation, most deep learning-based motion-related methods simulate motion in acquired images using retrospective transformation \cite{bmotionimage,bmotionimage1,bmotionimage2,bmotionimage3,bmotionksp}, hence, the accuracy is always limited by pixel size and cannot fulfill the demand in this work. For intra-shot motion, the degree of motion is often far less than the size of a pixel between different phase lines. Therefore, we adopted a different method for motion simulation, which applied the motion operator in the scanner coordinate system during the Bloch simulation. The results in Fig. 9 show that retrospective motion simulation methods can lead to signal corruption and signal loss, which further degrade the final motion-corrected T$_2$ maps. Though the Bloch-based simulation might not accurately reflect all possible forms of real artifacts, the results show that the artifacts are most successfully eliminated. 

Some novel unpaired learning algorithms have been published to overcome the lack of paired data in real world \cite{bganmoco,bganmoco2,byemoco}. Liu \emph{et al.} \cite{bganmoco} proposed a GAN-based framework to remove motion artifacts. They formulate the artifact removal problem as domain translation under the assumption that MR image is a nonlinear combination of content and artifact components. Though the paired data are not required, it is still necessary to manually distinguish between artifact-free and artifact-corrupt images to build a large realistic training dataset. Oh \emph{et al.} \cite{byemoco} converted motion artifact correction problem to subsampling MR reconstruction problem using the bootstrap subsampling and aggregation. However, as reported by the authors, this method faces challenges in intra-shot motion correction because the effect of intra-shot motion cannot be considered as sparse outliers in k-space. 

We believe that synthetic data-based approach may offer a new “unpaired learning” paradigm and can take full advantage of supervised learning. Moreover, model-based methods are often limited by the complexity of imaging models. The proposed MOST-DL provides a possible solution to simplify the model optimization problem in a data-driven manner, which can be used to address challenging topics in medical imaging.

\subsection{Non-ideal Factors Modeling and Reconstruction}
The modeling and reconstruction of non-ideal factors is a key feature with great potential in the MOST-DL framework. Combined with more complex encoding in the signal acquisition process (e.g., MOLED encoding), MOST-DL can achieve sophistication that was previously impossible. As shown in Fig. 11, we first present 2D rigid motion estimation at pixel level (velocity fields) of single-shot acquisition without any motion navigator. Motion information is often obtained from time series using image registration-based algorithms or tracking devices. For example, some approaches rely on motion-resolved imaging, which is achieved by modeling the signal correlation between different motion states along an additional motion-dedicated dimension \cite{b11}. However, these methods require the acquisition of a large number of time frames for a specific task. In contrast, with the help of the MOST-DL framework, we consider the motion estimation problem as a non-ideal factor reconstruction task, since the subject motion will bring extra phase accumulation and results in phase mismatch and artifacts. With paired synthetic data, the network is trained to learn motion pattern from motion-corrupted images with various levels of rigid motion and the results are mostly confirmed in \emph{in vivo} experiments. The reason may be that motion alters data distribution so that it can be distinguished by the neural network, as reported by Liu \emph{et al.} \cite{bganmoco}. Similarly, under the MOLED encoding and MOST-DL decoding, the B$_1^+$ field inhomogeneity can also be estimated, which provides a new way for B$_1^+$ mapping at high efficiency. 

In addition, non-ideal factor modeling and reconstruction may open a door to explore the domain gap between synthetic and real data. Specifically, during data generation, the MOST-DL framework allows the modeling of arbitrary new non-ideal factors to explore whether they affect the final results. Then, the non-ideal factors reconstruction provides a visual representation of the added non-ideal factors to validate the modeling plausibility. For example, in this work, the subject motion was modeled as a major non-ideal factor to generate training data for motion correction in T$_2$ mapping. The velocity fields estimation does not serve motion correction but provides a visualization of the instantaneous motion state, i.e., it explicitly indicates the motion information carried in the original data. By comparing with the reference velocity field, we have reason to believe that the motion modeling in the data generation is consistent with the real situation. 

\subsection{Extensions and Limitations}
The proposed method is not limited to the MOLED sequence and can be extended to other MRI pulse sequences and even other fields of model-based medical imaging. Expansion requires a full understanding of the physical model and consideration of the impact of various non-ideal factors. In principle, the generalizability of MOST-DL relies heavily on the versatility of the Bloch simulation in MRI signal evolution. For example, in inter-shot motion correction, a multi-shot pulse sequence (e.g., multi-shot EPI or TSE sequence) is needed for simulation with different motion patterns between shot to shot. The proposed Bloch-based motion modeling is still suitable for multi-shot acquisition and facilitates the correction of small subject motions at the sub-voxel level. Because it is beyond the scope of this article, the relevant results are not provided.

There are still several limitations of the proposed method. First, the public multi-contrast MRI datasets are not always sufficient in some specific anatomical regions such as the abdomen, prostate and knee. However, an increasing number of techniques have been proposed for missing MRI contrast synthesis. For example, Sharma \emph{et al.} \cite{bgan1} and Yurt \emph{et al.} \cite{bgan2} present frameworks to generate one or more missing contrasts by leveraging redundant information using GAN. These techniques could be applied to our proposed framework for relaxation parameters generation. Second, our method only simulates the in-plane rigid motion under the 2D MOLED acquisition, and severe through-plane motion still degrades the final results. Future work will focus on adapting the framework to 3D or non-rigid motion, which is increasingly used in clinical practice. Finally, the Bloch simulation used for data generation suffers from high computational costs even with GPU acceleration. A more efficient data generation technique is expected and will benefit our proposed supervised learning framework and reinforcement learning in medical imaging.

\section{Conclusion}
In this article, a model-based framework for synthetic data generation called MOST-DL was introduced. It was applied to solve a challenging problem of quantitative MRI under subject motion and non-ideal RF field. The results suggest that the MOST-DL method can generate synthetic images comparable to real data in quality, and achieve high performance in parallel reconstruction and motion correction. Since the proposed framework is based on Bloch simulation and general MRI models, we believe that it may be applicable to similar problems with other MRI acquisition methods.

\section{Acknowledgments}
The authors thank Fang Liu for developing the MRiLab software, and Michael Lustig, Jong Chul Ye for sharing their codes online. The authors sincerely appreciate the editor and reviewers for their valuable comments that helped us to significantly improve this work.

\end{document}